\documentclass[10pt,journal,compsoc]{IEEEtran}

\usepackage{url}
\usepackage{gensymb}
\usepackage{blindtext}
\usepackage{graphicx}
\usepackage{amsmath}
\usepackage{url}
\usepackage{subfig}
\usepackage{multirow}
\usepackage{hhline}
\usepackage{etoolbox}
\usepackage[english]{babel}
\usepackage{array}

\begin{document}

\title{Joint Image-Text News Topic Detection and Tracking with And-Or Graph Representation}

\author{Weixin~Li, Jungseock~Joo, Hang~Qi, and Song-Chun~Zhu
\IEEEcompsocitemizethanks{\IEEEcompsocthanksitem W. Li, J. Joo and H. Qi are with the Department of Computer Science, University of California, Los Angeles (UCLA). E-mail: \{lwx, joo, hangqi\}@cs.ucla.edu
\IEEEcompsocthanksitem S.-C. Zhu is with the Department of Statistics and Computer Science, UCLA. E-mail: sczhu@stat.ucla.edu}
}


\IEEEtitleabstractindextext{%
\begin{abstract}

In this paper, we aim to develop a method for automatically detecting and tracking topics in broadcast news. We present a hierarchical And-Or graph (AOG) to jointly represent the latent structure of both texts and visuals. The AOG embeds a context sensitive grammar that can describe the hierarchical composition of news topics by semantic elements about people involved, related places and what happened, and model contextual relationships between elements in the hierarchy. We detect news topics through a cluster sampling process which groups stories about closely related events. Swendsen-Wang Cuts (SWC), an effective cluster sampling algorithm, is adopted for traversing the solution space and obtaining optimal clustering solutions by maximizing a Bayesian posterior probability. Topics are tracked to deal with the continuously updated news streams. We generate topic trajectories to show how topics emerge, evolve and disappear over time. The experimental results show that our method can explicitly describe the textual and visual data in news videos and produce meaningful topic trajectories. Our method achieves superior performance compared to state-of-the-art methods on both a public dataset Reuters-21578 and a self-collected dataset named UCLA Broadcast News Dataset.

\end{abstract}

}

\maketitle

\IEEEdisplaynontitleabstractindextext

\IEEEpeerreviewmaketitle

\section{Introduction}

\subsection{Motivation and Objective} \label{motivation}

\IEEEPARstart{N}{ews} plays a vital role in informing citizens, affecting public opinions and policy making. The analyses of information flow in news media, such as selection and presentation biases, agenda-setting patterns, persuasion techniques, causal mechanisms, etc., are important issues in social and political science research. However, the sheer amount of news data overwhelms manual analysis. The objective of this paper is to develop an automatic topic detection and tracking method that provides a promising news parsing solution to serve as the basis for further analyses.





Detecting topics summarizes and organizes the large news collection, which contains rich textual and visual data. Both texts and visuals play key roles in the topic detection process. More importantly, they are both desired to be easily accessible to systematic research in social and political science. However, most of the traditional topic detection methods \cite{Allan98topicdetection, Blei:2012:PTM:2133806.2133826, LDA, dynamic_topic_model, text_clustering_survey, Xie} are single-modal and use texts only.


Moreover, for both texts and visuals, rather than utilizing the overall data, social and political scientists usually focus on some specific aspects when doing media analyses. Thus in contrast to generating coarse-grained topics (e.g. using the bag-of-words representation) on which most studies in the current literature concentrate \cite{LDA, Xie, Zhai:2005:TNS:1101149.1101152, 1621449}, they request accurate and fine-grained information in their analyses. For instance, in the bias analysis for the U.S. presidential election topic, the coverage of different candidates by different news networks is often compared \cite{SSQU:SSQU479, PSQ:PSQ2668}, which requires the extraction of candidate names from texts. A recent work used face images of candidates to predict the election outcomes \cite{Joo_face_traits}. Some other work also considered faces and scenes when studying the visual persuasion in election \cite{Joo_visual_persuasion, 10.2307/2111127}. Hence subcomponents from both texts and visuals, such as names and faces, are desired to be modeled and extracted in the topic detection process. In other words, instead of representing topics' texts and visuals at a coarse-grained level, it is preferred that finer-grained compositional topic representations can be used to provide a more detailed interpretation.



Tracking topics deals with the continuously updated news data. Our objective is to generate topic trajectories to show how topics emerge, evolve, and disappear, and how their subcomponents change over time. This is also demanded by a number of applications such as the causal mechanism analysis \cite{doi:10.1080/10810730305724}. Traditional topic tracking which is defined as the process of tracking the recurrence of known topics in new incoming stories \cite{TDT_book, Allan98topicdetection} thus can hardly fulfill this goal. Some other methods, e.g. \cite{dynamic_topic_model}, can model topic over time but fail to efficiently deal with the updated news data.

Despite the decades of study, there lacks publicly available multimedia datasets for evaluating news topic detection and tracking methods. Even though some multimedia news datasets have been used in previous work, such as the TDT datasets \cite{TDT_book}, and the TRECVID corpus \cite{1178722}, they are not publicly available, and some of them do not have ground-truth annotations.


To solve the aforementioned problems, in this paper, we present a method for joint image-text news topic detection and tracking.  Both texts and visuals, along with their subcomponents are modeled using a compositional topic representation. For evaluation, we use data from the UCLA Library Broadcast NewsScape\footnote{http://newsscape.library.ucla.edu}, which contains a large number of broadcast news programs from U.S. and around the world since 2005. To collect the ground-truth data, we annotate a subset from the large collection. 


We also made case studies based on our method, including tracking the 2016 U.S. presidential election and analyzing the gun shooting events. We have built a website Viz2016 to visualize our large-scale election tracking results\footnote{http://viz2016.com}. The results for gun shooting events will be shown in the experiment part of this paper.

\begin{figure}
\centering
\includegraphics[width=3.3in]{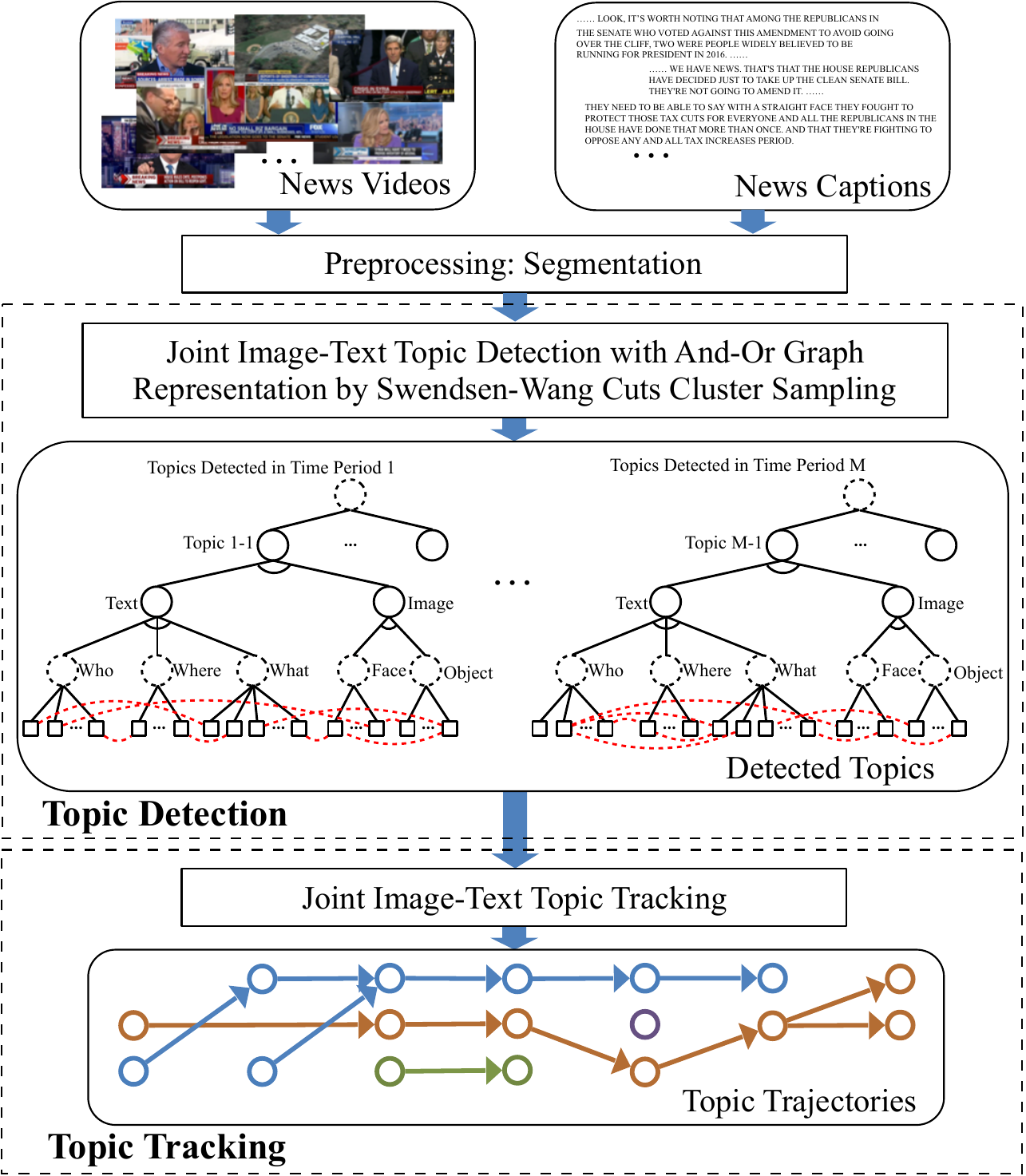}
\caption{Overview of the proposed topic detection and tracking method. The inputs include both news videos and closed captions (texts). We detect topics through a joint image-text cluster sampling method within different time periods. Then the detected topics are tracked over time to form topic trajectories.}
\label{framework_fig}
\end{figure}

\subsection{Overview of Our Method}

Fig. \ref{framework_fig} is an overview of our topic detection and tracking method. Both news videos and closed captions (texts) are included as the input of our method. After preprocessing the input including steps such as story segmentation, we detect topics using a cluster sampling method based on the And-Or graph (AOG) topic representation which jointly models texts and images in news videos and organizes news topic components in a hierarchical structure. We further link topics detected in different time periods to generate topic trajectories that show how topics evolve over time.

\begin{figure*}
\centering
\includegraphics[width=6.6in]{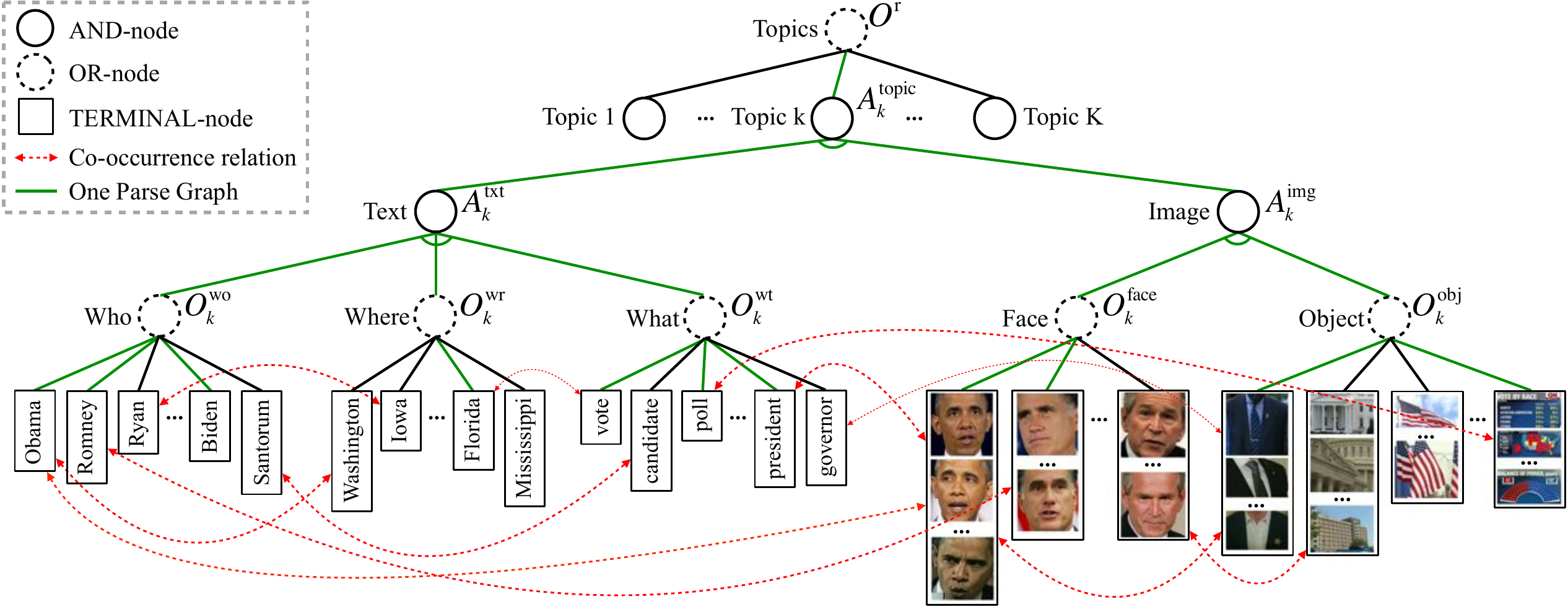}
\caption{Illustration of the And-Or graph topic representation. Three types of nodes are included: AND-nodes representing topics' compositions (e.g. a topic is composed of the text part and the image part), OR-nodes for alternative structures (e.g. different configurations of a component in the topic structure), and TERMINAL-nodes representing the most elementary components. The dashed red lines represent different components' co-occurring pairs. The green lines show an example of the parse graph.}
\label{AOG_fig}
\end{figure*}

\subsubsection{The And-Or Graph Representation}

The proposed AOG embeds a context-sensitive grammar that jointly models hierarchical topic compositions of texts and images. Fig. \ref{AOG_fig} illustrates the AOG topic representation:
\begin{itemize}
\item \textit{The root OR-node} $O^{\rm{r}}$ in the top layer represents different topic configurations.
\item Each topic configuration is then represented by \textit{a single topic AND-node} $A_k^{\rm{topic}}$ ($k = 1, ..., K$ where $K$ is the total topic number) in the second layer. This node is composed of two parts, which represent texts and images respectively.
\end{itemize}


\textbf{Text Representation}. The text part of each topic is represented by an \textit{AND-node}, as shown in Fig. \ref{AOG_fig} (node $A_k^{\rm{txt}}$). This node has three components, which encode the knowledge of "who", "where" and "what". These three components describe the people involved, related places, and what happened respectively, which are three major aspects of new events and topics.

The "who", "where" and "what" components are all represented by \textit{OR-nodes} (nodes $O_k^{\rm{wo}}$, $O_k^{\rm{wr}}$, and $O_k^{\rm{wt}}$ in Fig. \ref{AOG_fig}). All of these nodes can describe a set of possible words for the corresponding components. A certain news story may trigger a subset of these words. The words are represented by \textit{TERMINAL-nodes} in the last layer. We also embed the contextual relations between the "who", "where" and "what" components in the AOG. They are described using information from two aspects: 
\begin{itemize}
\item the co-occurrences of words from different components (such as the co-occurring pairs marked by the dashed red lines in Fig. \ref{AOG_fig});
\item the ratios of entity numbers of different components (e.g. some topics have more people involved compared to the related locations).
\end{itemize}

\textbf{Image Representation}. The image part of each topic is also represented by an \textit{AND-node} as shown in Fig. \ref{AOG_fig} (node $A_k^{\rm{img}}$). This node has two components, which capture two important visual signals in news: faces and objects. Faces show the main people related to the topic, and objects include other general information about the scene and the event.

The face and object components are represented by \textit{OR-nodes} (nodes $O_k^{\rm{face}}$ and $O_k^{\rm{obj}}$ in Fig. \ref{AOG_fig}), which can describe a set of possible entities similar to the "who", "where", and "what" OR-nodes. Each face/object entity corresponds to one cluster of face/object patches, and we use a \textit{TERMINAL-node} to represent it in the last layer. We also encode the contextual relations between the face and object components in the AOG using co-occurrences of face-object pairs (such as the co-occurrence of politicians and suits).

\textbf{Joint Image-Text Representation}. To jointly model the topic text and image parts, we also describe their contextual relations in the AOG. The ratio of the total entity numbers of these two parts (i.e. numbers of all TERMINAL-nodes under the nodes $A_k^{\rm{txt}}$ and $A_k^{\rm{img}}$ in Fig. \ref{AOG_fig}) is included in the AOG. We also model three component pair relations selected from these two parts, namely face-who, face-what and object-what pairs. The face-who and face-what pairs can clearly relate the faces appeared in the video to their names and other co-appearing textual knowledge respectively. The object-what pairs can relate the objects to textual descriptions. For these three component pairs, the pair instance frequencies are used to model the contextual relations.

In summary, the proposed AOG topic representation jointly models texts and images and their subcomponents in a hierarchical structure. The AOG model strikes a balance between the syntactic representation in NLP (too complex to compute) and the simplistic BoW representation (too coarse). It  supports the news topic detection and tracking tasks with the appropriate complexity accurately.

\subsubsection{Task: Detecting and Tracking News Topics}

In the massive and continuously updated news data, each news topic evolves over time. We aim to detect topics within short time periods and further generate long-time topic trajectories. Therefore, we can show both detailed descriptions for each topic in different time periods, and how each topic develops over time. It also helps prevent the heavy computation incurred by periodically detecting topics using the entire updated news collection.

For topic detection, we group stories that elaborate the same topics. The proposed AOG explicitly describes components of different topics. Thus based on the AOG, we can effectively group related stories and generate meaningful topics. We solve the grouping problem by cluster sampling methods by maximizing a Bayesian posterior probability. An efficient cluster sampling algorithm introduced in image segmentation, i.e. Swendsen-Wang Cuts (SWC) \cite{swc_paper}, is adopted for topic detection.

For topic tracking, with the hierarchical AOG model that can represent topic compositions and how such information changes over time, we align news topics in different time periods. This provides us a promising way to track and keep updating the news states. We link topics detected in different short time periods to generate topic trajectories by considering both topic similarities and their temporal relations.

In the experiments, we show that our method can generate meaningful topics and topic trajectories. It also achieves better performance compared to state-of-the-art algorithms.

\subsection{Summary of Contributions}

This paper makes the following contributions:
\begin{itemize}
\item We proposed a joint image-text compositional news topic representation based on And-Or Graph, which better utilize the multimodal data and provide interpretations with appropriate amount of details compared to single-modal methods and other representations \cite{Allan98topicdetection, LDA, Xie, Zhai:2005:TNS:1101149.1101152, 1621449, Griffiths05integratingtopics, Boyd-graber_syntactictopic}. 
\item We solve the topic detection problem using the clustering sampling method Swendsen-Wang Cuts, which has better performance than commonly used greedy algorithms \cite{LDA, Xie, 1621449, 4432629}.
\item We detect and track topics simultaneously over time, generating both topic summaries in different time periods and long-time topic trajectories. The results show how topics evolve over time and provide useful data for further media analysis, which can hardly be fulfilled by traditional topic detection methods \cite{TDT_book, Allan98topicdetection}.
\item We collected a news dataset for joint image-text topic detection and tracking, and also provide the ground-truth annotations, which copes with the lack of publicly available multimodal news datasets.
\end{itemize}

The rest of the paper is organized as follows. We first review the related literature in Section \ref{related_work}. Then we present the proposed topic representation in Section \ref{topic_representation}. The proposed topic detection method is described in Section \ref{topic_detection}. The topic tracking method is illustrated in Section \ref{topic_tracking}. We report our experiment results and comparisons with other state-of-the-art methods in Section \ref{experiments} and conclude in Section \ref{conclusion}.

\section{Related Work} \label{related_work}

Our work is mainly related to the following four research streams: (1) topic modeling, (2) topic clustering, (3) topic tracking, and (4) news gathering and delivering systems.

\subsection{Topic Modeling}

Among the large number of topic modeling methods, probabilistic topics models \cite{Hofmann:1999:PLS:312624.312649, Blei:2012:PTM:2133806.2133826} have been effectively used for detecting and analyzing latent topics, such as the latent Dirichlet allocation (LDA) model \cite{LDA, griffiths_steyvers04} and its extensions \cite{Teh04hierarchicaldirichlet, Wang:2011:CTM:2020408.2020480, blei2007correlated, dynamic_topic_model, Wallach:2006:TMB:1143844.1143967, Wang:2006:TOT:1150402.1150450}. 

Even though these methods are effective in general topic modeling, they can hardly achieve good performance in the news domain using only the bag-of-words (BoW) representation. The BoW representation is computationally efficient, but it ignores the compositional structures, which are important for news analyses. News stories are generally driven by events, so information from aspects like "who", "where" and "what" is crucial for summarizing these stories and generating meaningful news topics. Newman et al. \cite{Newman:2006:SEM:1150402.1150487} considered these aspects but included them as a whole. Li et al. \cite{Li:2005:PMR:1076034.1076055} used information from the above aspects in their representation. However, they assume that these aspects are independent, which is generally not true in the real news data.

Moreover, all the aforementioned topic modeling methods are single-modal methods which only use texts. Several multi-modal probabilistic topic models have been proposed for other tasks such as image annotation and classification \cite{Blei:Modeling_annotated_data, 5540000, Niu:2014:SRT:2679600.2680028}, news geo-location inference \cite{Zhou:2012:GIN:2393347.2396301}, etc., but methods for joint image-text topic detection are rare in the literature.

\subsection{Topic Clustering}

Clustering based methods are also widely used for the task of news topic detection. They assume that each news story talks about one news event which corresponds to one topic. A large number of methods for topic detection in the Topic Detection and Tracking (TDT) research \cite{TDT_book} (e.g. \cite{Allan98topicdetection, Yamron99topictracking}) use clustering methods for detecting news topics, where stories on the same topic are gathered. Traditional document clustering methods \cite{text_clustering_survey, Steinbach00acomparison} can also be used for topic detection. However, most of these methods are single-modal and mainly focus in the text domain.

Multimodal topic clustering methods have been proposed by taking both texts and visuals into consideration. In most of these methods, texts are represented using the BoW representation \cite{1621449, 4432629, 6877623, Zhai:2005:TNS:1101149.1101152}. For visual representation, some methods use color histograms of the keyframes \cite{1621449}. Other methods detect the near-duplicate keyframes (NDK) first and then use them to build visual relations between news stories \cite{4432629, 6877623}. Even though these methods can compute the visual similarities between stories, they are not capable of modeling the visual part decomposition of news topics. In terms of the clustering methods, \cite{1621449} and \cite{4432629} used co-clustering algorithm and one of its extensions with constraints added respectively. \cite{Zhai:2005:TNS:1101149.1101152} groups news stories based on the linear combination of textual and visual similarities. \cite{6877623} detects topics within one multi-modality graph, which is obtained by merging one text graph and another visual graph constructed based on LDA and NDK respectively. These clustering methods are not global optimal.

Some work also combined topic modeling and document clustering together, such as the multi-grain clustering topic model (MGCTM) proposed by Xie et al. \cite{Xie}. They showed that these two tasks are closely related and can help each other as both performances are improved. This work still remains in the pure text domain and uses the BoW representation.

\subsection{Topic Tracking}


The traditional topic tracking problem in TDT \cite{TDT_book, Allan98topicdetection} is defined as the process of finding related additional stories for some pre-learned topics. Many methods have been proposed for solving this problem such as those in \cite{TDT_book, Makkonen:2004:SST:964569.964611, 4106486}. However, deciding the topic of each new-coming story based on the previous learned topics can take a long time in a large data collection. In addition, they can hardly deal with the newly emerging topics.

In the probabilistic modeling community, some models incorporate time information, such as the Dynamic Topic Model (DTM)\cite{dynamic_topic_model} which can model the topic evolution over time. However, it is assumed that the topics exist throughout the whole time period, which is usually not true especially in broadcast news. It also leads to heavy computation for continuously updated new streams.

Thus instead of using the previous two methods, we choose to track the news topics by linking topics detected in different time periods and generating topic trajectories over time. Some linking methods, such as those by Mei and Zhai \cite{Mei:2005:DET:1081870.1081895} as well as Kim and Oh \cite{Kim:2011:TCU:1964750.1964765}, are closely related to our topic tracking task. However, the method in \cite{Mei:2005:DET:1081870.1081895} is designed for news about some specific topics such as "tsunami". The similarity matrices used in \cite{Kim:2011:TCU:1964750.1964765} are based on the topics obtained by the original LDA model with the BoW assumption. Moreover, both of these two methods are merely based on textual information.

\subsection{News Gathering and Delivering System}

Several news gathering and delivering systems have been presented recently. In \cite{Jou:2013:SEH:2502081.2508118, newsrover}, Chang et al. presented a system called News Rover to integrate multimodal news sources. News topics are collected from sources such as Google News and organized in a hierarchical topic structure, and news stories are then matched to these topics. Another personalized news video system, EigenNews \cite{conf/mm/YuVCTDACG13, 6618439}, can aggregate news videos from multiple sources. It matches the extracted news stories to online news articles to get the related news categories. In this paper, we detect and track news topics in a fully unsupervised way by jointly modeling texts and visuals to support the media analyses in social and political science (e.g. the two case studies mentioned in Section \ref{motivation}).


\section{Topic Representation} \label{topic_representation}

In this section, we define the And-Or graph (AOG) for topic representation.

\subsection{Overall Representation}

An \textbf{AOG} embodies a context sensitive grammar. It can be defined by a three-tuple $\mathcal G = (V, E, \Theta)$. The node set $V$ consists of three subsets of nodes: \textbf{AND-nodes} $V_{\rm{AND}}$, \textbf{OR-nodes} $V_{\rm{OR}}$ and \textbf{TERMINAL-nodes} $V_{\rm{T}}$, i.e. $V = V_{\rm{AND}} \cup V_{\rm{OR}} \cup V_{\rm{T}}$. $E$ denotes the edge set in the graph. $\Theta$ represents the AOG model parameters. We have $\Theta = \{K, \theta_1, ..., \theta_K\}$ where $K$ is the total topic number, and $\theta_1, ..., \theta_K$ represent the model parameters for these $K$ topics respectively. Fig. \ref{AOG_fig} illustrates a small part of the proposed AOG topic representation.


A \textbf{parse graph} $pg$ is an instantiation of the AOG by selecting children nodes at OR-nodes. The green lines in Fig. \ref{AOG_fig} shows one example of the parse graph.



As shown in Fig. \ref{AOG_fig}, the AOG has five layers. Nodes in each layer are explained as follows:

1) \textbf{The root OR-node} $O^{\rm{r}} \in V_{\rm{OR}}$ in the first layer of the AOG represents different topic configurations and their mutual contextual information. Each topic $k$ ($k = 1, ..., K$) is represented by an AND-node $A_k^{\rm{topic}}$ in the second layer of the AOG hierarchy with the model parameter $\theta_k$. 


News stories are reports of topics, i.e. topic instances, from various TV news networks. We denote a news story by ${\bf{d}}_i$. For a story ${\bf{d}}_i$, the scoring function at the root OR-node $O^{\rm{r}}$ is defined as:
\begin{equation}
{score}^{\rm{root}}({\bf{d}}_i; \Theta) = \mathop {\max}\limits_{\theta_k \in \Theta} {score}^{\rm{topic}}({\bf{d}}_i; \theta_k),
\end{equation}
where $ {score}^{\rm{topic}}({\bf{d}}_i; \theta_k)$ is the scoring function at $A_k^{\rm{topic}}$, which will be introduced later.

2) \textbf{The topic AND-node} $A_k^{\rm{topic}} \in V_{\rm{AND}}$ in the second layer of the hierarchy (as shown in Fig. \ref{AOG_fig}) represents one topic configuration. One topic is composed of the text part and the image part. So $A_k^{\rm{topic}}$ has two children AND-nodes, i.e. the text AND-node $A_k^{\rm{txt}}$ and the image AND-node $A_k^{\rm{img}}$. The scoring function at $A_k^{\rm{topic}}$ is defined as:
\begin{equation}
\begin{aligned}
{score}^{\rm{topic}}({\bf{d}}_i; \theta_k) = {score}^{\rm{txt}}({\bf{d}}_i^{\rm{txt}}; \theta_k) + {score}^{\rm{img}}({\bf{d}}_i^{\rm{img}}; \theta_k)\\ + {score}^{\rm{joint}}({\bf{d}}_i^{\rm{joint}}; \theta_k) + g(f_{A_k^{\rm{topic}}}),
\end{aligned}
\label{eq_topic_and}
\end{equation}
where ${\bf{d}}_i^{\rm{txt}}$, ${\bf{d}}_i^{\rm{img}}$ and ${\bf{d}}_i^{\rm{joint}}$ denote the text part, the image part and their joint information of the story ${\bf{d}}_i$ respectively (${\bf{d}}_i = {\bf{d}}_i^{\rm{txt}} \cup {\bf{d}}_i^{\rm{img}} \cup {\bf{d}}_i^{\rm{joint}}$). The two terms ${score}^{\rm{txt}}({\bf{d}}_i^{\rm{txt}}; \theta_k)$ and ${score}^{\rm{img}}({\bf{d}}_i^{\rm{img}}; \theta_k)$ are the scoring functions at $A_k^{\rm{txt}}$ and $A_k^{\rm{img}}$ respectively. The term ${score}^{\rm{joint}}({\bf{d}}_i^{\rm{joint}}; \theta_k)$ describes the contextual relations between the text part and the image part. These three terms will be explained later. The function $g(f_{A_k^{\rm{topic}}})$ describes the prior of choosing $A_k^{\rm{topic}}$ at root node $O^{\rm{r}}$. We have the branching frequency $f_{A_k^{\rm{topic}}} \in \theta_k$. We observed that in the broadcast news, dominant topics with a large amount of coverage only constitute a small portion of the whole corpus, and the sizes of most topics are small. Accordingly, we assume that the branching frequencies at $O^{\rm{r}}$ follow a power law distribution\footnote{In the experiments, for the function $g(\cdot)$, we use the Zipf's law probability distribution, i.e. $g(f) = \frac{{f^{-s}}}{{\zeta(s)}}$ and set the parameter $s$ that describes the distribution's exponent as $s = 1.75$ ($\zeta$ is the Riemann Zeta function).} (the verification of our observation will be shown in Section \ref{observation_verification}).

\subsection{Text Representation}

For a news story ${\bf{d}}_i$, its \textit{text part} ${\bf{d}}_i^{\rm{txt}}$ contains the "who" component ${\bf{d}}_i^{\rm{wo}}$, the "where" component ${\bf{d}}_i^{\rm{wr}}$, and the "what" component ${\bf{d}}_i^{\rm{wt}}$. We extract words for different components by performing the name entity extraction using the Stanford Named Entity Recognizer \cite{Finkel:2005:INI:1219840.1219885}. Thus each of the three components can be represented by a list of words (word duplication is allowed in the list), e.g. ${\bf{d}}_i^{\rm{wo}} = (w_1, ..., w_{M_i^{\rm{wo}}})$ where ${M_i^{\rm{wo}}}$ is the total number of words in the "who" component in the story ${\bf{d}}_i$. The total numbers of words in the "where" and "what" components are denoted by $M_i^{\rm{wr}}$ and $M_i^{\rm{wt}}$ respectively.


We extract the co-occurring word pairs from the three components in the text part. We consider a pair of words as one co-occurring pair if the two words belong to two different components, and are extracted from the same sentence. The list of co-occurring word pairs in the story ${\bf{d}}_i$ is denoted by ${\bf{d}}_i^{\rm{tt}} = [(w_1, w_2) | w_1 \in {\bf{d}}_i^{\rm{wo}}, w_2 \in {\bf{d}}_i^{\rm{wr}}] \cup [(w_1, w_2) | w_1 \in {\bf{d}}_i^{\rm{wo}}, w_2 \in {\bf{d}}_i^{\rm{wt}}] \cup [(w_1, w_2) | w_1 \in {\bf{d}}_i^{\rm{wr}}, w_2 \in {\bf{d}}_i^{\rm{wt}}]$.


\textbf{The text AND-node} $A_k^{\rm{txt}}$ in the third layer of the AOG hierarchy has three children OR-nodes, i.e. $O_k^{\rm{wo}}$, $O_k^{\rm{wr}}$, and $O_k^{\rm{wt}}$, which represent the "who", "where" and "what" components in the text part of topic $k$ respectively. The scoring function at $A_k^{\rm{txt}}$ (i.e. the term ${score}^{\rm{txt}}({\bf{d}}_i^{\rm{txt}}; \theta_k)$ in Eq. \ref{eq_topic_and}) is defined as:
\begin{equation}
\begin{aligned}
{score}^{\rm{txt}}({\bf{d}}_i^{\rm{txt}}; \theta_k) = \sum\limits_{\substack{c}} {{score}^{\rm{comp}}({\bf{d}}_i^{c}; \theta_k)}\\ + {score}^{\rm{tt}}({\bf{d}}_i^{\rm{tt}}; \theta_k),
\end{aligned}
\end{equation}
where the variable $c$ represents the component type $c \in \{{\rm{wo}}, {\rm{wr}}, {\rm{wt}}\}$. The term ${{score}^{\rm{comp}}({\bf{d}}_i^{c}; \theta_k)}$ represents the scoring function at the OR-node for one component $O_k^{c} \in \{O_k^{\rm{wo}}, O_k^{\rm{wr}}, O_k^{\rm{wt}}\}$. The term ${score}^{\rm{tt}}({\bf{d}}_i^{\rm{tt}}; \theta_k)$ describes the contextual relations between the three components in the text part and we define it as:
\begin{equation}
\begin{aligned}
{score}^{\rm{tt}}({\bf{d}}_i^{\rm{tt}}; \theta_k) = \sum\limits_{c_1, c_2} G(\frac{{M_i^{c_1}}}{M_i^{c_2}} ; \mu_k^{{c_1}{c_2}}, \sigma_k^{{c_1}{c_2}}) \\ + \sum\limits_{(w_1, w_2) \in {\bf{d}}^{\rm{tt}}} \log({f}_k^{(w_1, w_2)} + 1),
\end{aligned}
\label{text_contexts}
\end{equation}
where we have the component types $c_1 \in \{\rm{wo}, \rm{wr}\}, c_2 \in \{\rm{wr}, \rm{wt}\}, c_1 \neq c_2$. $\frac{{M_i^{c_1}}}{M_i^{c_2}}$ represents the ratio of word numbers from two different components. The three ratios, namely $\frac{{M_i^{\rm{wo}}}}{M_i^{\rm{wr}}}$, $\frac{{M_i^{\rm{wo}}}}{M_i^{\rm{wt}}}$, and $\frac{{M_i^{\rm{wr}}}}{M_i^{\rm{wt}}}$ are assumed to follow Gaussian distributions. $\mu_k^{{c_1}{c_2}}, \sigma_k^{{c_1}{c_2}} \in \theta_k$ are the parameters for the corresponding Gaussian distributions. The verification of our assumption will be discussed in Section \ref{observation_verification}. The parameter ${f}_k^{(w_1, w_2)} \in \theta_k$ is the frequency of the co-occurring word pair $(w_1, w_2)$ in topic $k$.

\textbf{The three children OR-nodes of $A_k^{\rm{txt}}$} in the fourth layer, namely $O_k^{\rm{wo}}$, $O_k^{\rm{wr}}$, and $O_k^{\rm{wt}}$, describe a set of possible words for the corresponding components. The words are represented by TERMINAL-nodes in the last layer. The scoring functions at these OR-nodes are defined as:
\begin{equation}
{score}^{\rm{comp}}({\bf{d}}_i^{c}; \theta_k) = \sum\limits_{w \in {\bf{d}}_i^{c}} \log ({f}_k^w + 1)
\label{eq_or_fourth_layer}
\end{equation}
for component type $c \in \{{\rm{wo}}, {\rm{wr}}, {\rm{wt}}\}$. The parameter ${f}_k^w \in \theta_k$ represents the frequency of word $w$ in topic $k$.

\subsection{Image Representation}

\begin{figure}
\centering
\includegraphics[width=2.5in]{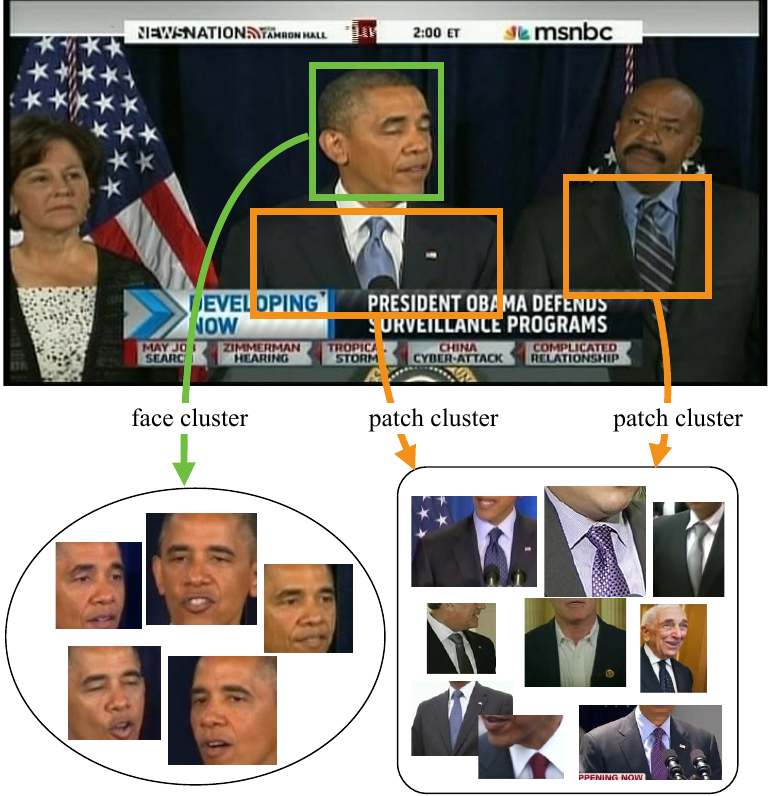}
\caption{Illustration of how one image can be parsed based on face and object clusters.}
\label{example_img_parse}
\end{figure}

The story's \textit{image part} ${\bf{d}}_i^{\rm{img}}$ contains the face component ${\bf{d}}_i^{\rm{face}}$, and the object component ${\bf{d}}_i^{\rm{obj}}$, i.e. ${\bf{d}}_i^{\rm{face}}, {\bf{d}}_i^{\rm{obj}} \in {\bf{d}}_i^{\rm{img}}$. Each entity in the face/object component corresponds to one cluster of face/object patches. To obtain the face component, we first perform face detection and extract face features based on Local Binary Pattern \cite{1717463} and Local Gabor Binary Pattern Histogram Sequence \cite{1541333}, and then use the k-means algorithm to cluster the faces into groups. To get the object component, we first extract patches from images using Selective Search \cite{UijlingsIJCV2013} which can generate possible object locations. We then get the patch features using Caffe \cite{jia2014caffe}, an open-source implementation of the deep convolutional network that is trained on over a million images annotated with 1,000 ImageNet \cite{ILSVRC15} classes. Then the k-means algorithm is used to cluster the patches into groups. Fig. \ref{example_img_parse} illustrates how one image can be parsed based on the obtained face and object clusters. Each face/object patch can be represented by its corresponding cluster membership. Then the face and object components of one story ${\bf{d}}_i$ can also be represented by a list of visual words, e.g. ${\bf{d}}_i^{\rm{face}} = (w_1, ..., w_{M_i^{\rm{face}}})$ where each word $w_j \in {\bf{d}}_i^{\rm{face}}$ represent one face patch's cluster membership. $M_i^{\rm{face}}$ is the total number of face patches in ${\bf{d}}_i^{\rm{img}}$ and the total number of object patches is denoted by $M_i^{\rm{obj}}$.

We extract the co-occurring word pairs from the face and object components of the image part. A pair of visual words is considered as one co-occurring pair if the two words are from the face and object components respectively, and they both appear in one short time period in the news video. We denote the list of co-occurring pairs extracted from the image part by ${\bf{d}}_i^{\rm{ii}} = [(w_1, w_2) | w_1 \in {\bf{d}}_i^{\rm{face}}, w_2 \in {\bf{d}}_i^{\rm{obj}}]$.

\textbf{The image AND-node} $A_k^{\rm{img}}$ in the third layer of the AOG represents the image part of the topic $k$. It has two children OR-nodes, i.e. $O_k^{\rm{face}}$ and $O_k^{\rm{obj}}$, which represent the face and object information respectively. The scoring function at $A_k^{\rm{img}}$ is defined in a similar way to the one at $A_k^{\rm{txt}}$. We have:
\begin{equation}
\begin{aligned}
{score}^{\rm{img}}({\bf{d}}_i^{\rm{img}}; \theta_k) = \sum\limits_{c} {{score}^{\rm{comp}}({\bf{d}}_i^{c}; \theta_k)}\\ + {score}^{\rm{ii}}({\bf{d}}_i^{\rm{ii}}; \theta_k),
\end{aligned}
\end{equation}
where the component type $c \in \{{\rm{face}}, {\rm{obj}}\}$. The term ${{score}^{\rm{comp}}({\bf{d}}_i^{c}; \theta_k)}$ represents the scoring function at the OR-node for one component $O_k^{c} \in \{O_k^{\rm{face}}, O_k^{\rm{obj}}\}$.

The term ${score}^{\rm{ii}}({\bf{d}}_i^{\rm{ii}}; \theta_k)$ describes the contextual relations between the face and object components and we define it as:
\begin{equation}
\begin{aligned}
{score}^{\rm{ii}}({\bf{d}}_i^{\rm{ii}}; \theta_k) = \sum\limits_{(w_1, w_2) \in {\bf{d}}^{\rm{ii}}} \log({f}_k^{(w_1, w_2)} + 1),
\end{aligned}
\end{equation}
where ${f}_k^{(w_1, w_2)} \in \theta_k$ is the frequency for the co-occurring visual word pair $(w_1, w_2)$ in the topic $k$.

\textbf{The two children OR-nodes of $A_k^{\rm{img}}$} in the fourth layer, namely $O_k^{\rm{face}}$, and $O_k^{\rm{obj}}$, can describe a set of alternative visual words. These words are represented by TERMINAL-nodes in the last layer. The scoring functions at these OR-nodes, i.e. ${{score}^{\rm{comp}}({\bf{d}}_i^{c}; \theta_k)}, c \in \{\rm{face}, \rm{obj}\}$, are defined in the same way as those at $O_k^{\rm{wo}}$, $O_k^{\rm{wr}}$ and $O_k^{\rm{wt}}$ (Eq. \ref{eq_or_fourth_layer}).

\subsection{Joint Textual and Visual Information Representation}

To jointly model the textual and visual information, we extract the co-occurring word pairs from the story's image and text parts. Three kinds of pairs, namely the face-who, face-what, and object-what pairs, are obtained for each news story ${\bf{d}}_i$. The words in each co-occurring pair appear in one short time period. These text-image co-occurring word pairs are denoted by ${\bf{d}}_i^{\rm{joint}} = [(w_1, w_2) | w_1 \in {\bf{d}}_i^{\rm{wo}}, w_2 \in {\bf{d}}_i^{\rm{face}}] \cup [(w_1, w_2) | w_1 \in {\bf{d}}_i^{\rm{wt}}, w_2 \in {\bf{d}}_i^{\rm{face}}] \cup [(w_1, w_2) | w_1 \in {\bf{d}}_i^{\rm{wt}}, w_2 \in {\bf{d}}_i^{\rm{obj}}]$.


The term ${score}^{\rm{joint}}({\bf{d}}_i^{\rm{joint}}; \theta_k)$ in Eq. \ref{eq_topic_and} describes the contextual relations between the text part and the image part. We use ${M_i^{{\rm{text}}}}$ and ${M_i^{{\rm{img}}}}$ to denote the total entity numbers of the text part and the image part respectively. So we have ${M_i^{{\rm{text}}}} = {M_i^{{\rm{wo}}}} + {M_i^{{\rm{wr}}}} + {M_i^{{\rm{wt}}}}$, and ${M_i^{{\rm{img}}}} = {M_i^{{\rm{face}}}} + {M_i^{{\rm{obj}}}}$. The score function ${score}^{\rm{joint}}({\bf{d}}_i^{\rm{joint}}; \theta_k)$ is defined as:
\begin{equation}
\begin{aligned}
{score}^{\rm{joint}}({\bf{d}}_i^{\rm{joint}}; \theta_k) = 
G(\frac{{M_i^{{\rm{text}}}}}{{M_i^{{\rm{img}}}}}; \mu_k^{\rm{joint}}, \sigma_k^{\rm{joint}}) \\ + \sum\limits_{(w_1, w_2) \in {\bf{d}}^{\rm{joint}}} \log({f}_k^{(w_1, w_2)} + 1).
\end{aligned}
\label{eq_txt_img}
\end{equation}
We assume that the ratio between the total entity numbers of the text part and the image part, i.e. $\frac{{M_i^{{\rm{text}}}}}{{M_i^{{\rm{img}}}}}$, follows Gaussian distribution with the parameters $\mu_k^{\rm{joint}}, \sigma_k^{\rm{joint}} \in \theta_k$. The verification of our assumption will be shown in Section \ref{observation_verification}. The parameter ${f}_k^{(w_1, w_2)} \in \theta_k$ is the frequency of the word pair $(w_1, w_2) \in {\bf{d}}^{\rm{joint}}$ in topic $k$.


Based on the previous scoring functions, we can find the optimal parse graph $pg^{*}_i$ for the story ${\bf{d}}_i$ by calculating ${score}^{\rm{root}}({\bf{d}}_i; \Theta)$.

\subsection{Empirical Evaluations of Assumptions in AOG} \label{observation_verification}

In the AOG representation, we make some assumptions of the distribution of the branching frequencies at $O^{\rm{r}}$ and the ratios between different components. To verify our assumptions, we collect a news corpus that contains news data during a period of seven days. There are 1,853 news stories in the corpus. Annotators are asked to group the stories according to their topics. After annotation, we got 355 topics in total.

\begin{figure}
\centering
\includegraphics[width=3in]{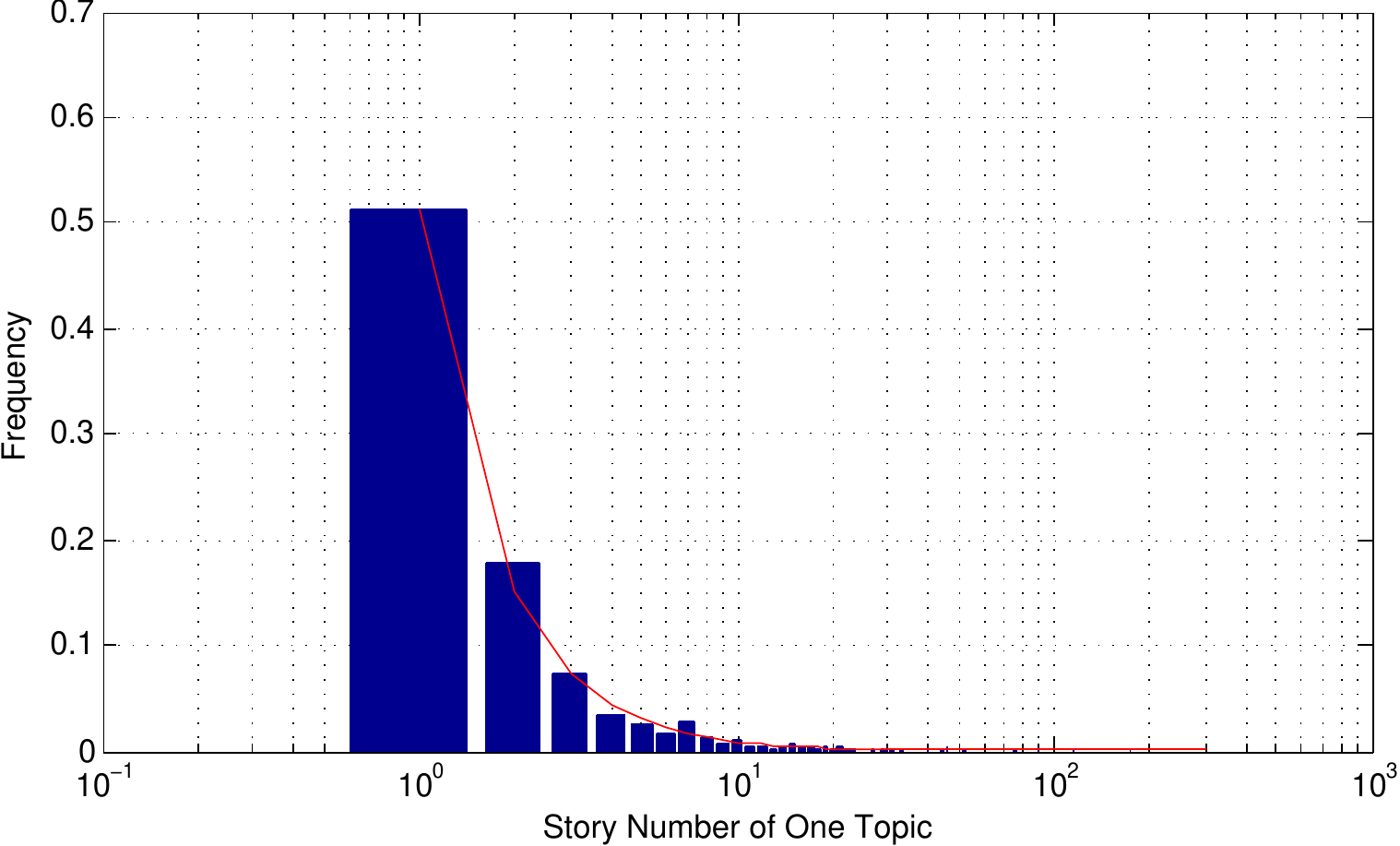}
\caption{Empirical histogram of the topic's story number and the fitting result (the red curve).}
\label{power_law_plot}
\end{figure}

To verify the assumption that the branching frequencies at $O^{\rm{r}}$ follow the power law distribution, using the collected corpus, we fit the empirical distribution of the story numbers in the topics (i.e. the topic branching frequency) to the power law distribution. The p-value (at the 5\% significance level) is 0.9984. Fig. \ref{power_law_plot} shows the empirical distribution and the fitted curve (red line).

\begin{figure}[]
\centering
\subfloat[]{\includegraphics[height=0.8in]{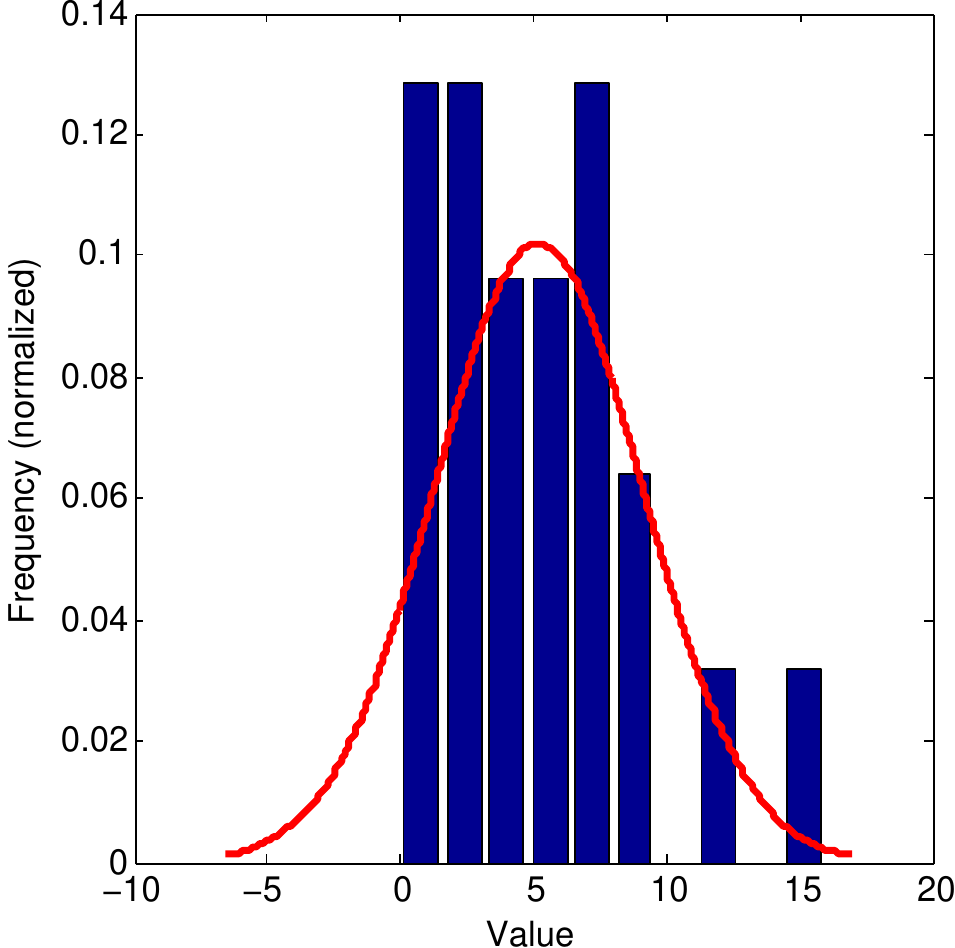}
\label{gau_who_where}}
\subfloat[]{\includegraphics[height=0.8in]{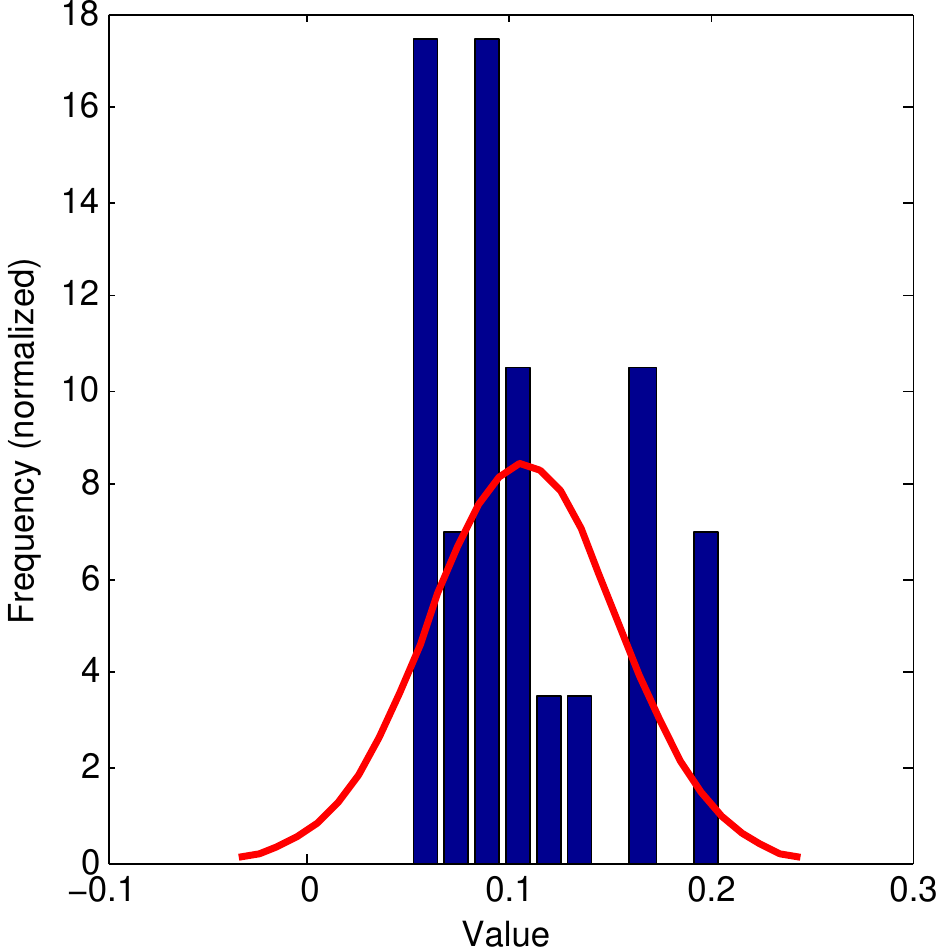}
\label{gau_who_what}}
\subfloat[]{\includegraphics[height=0.8in]{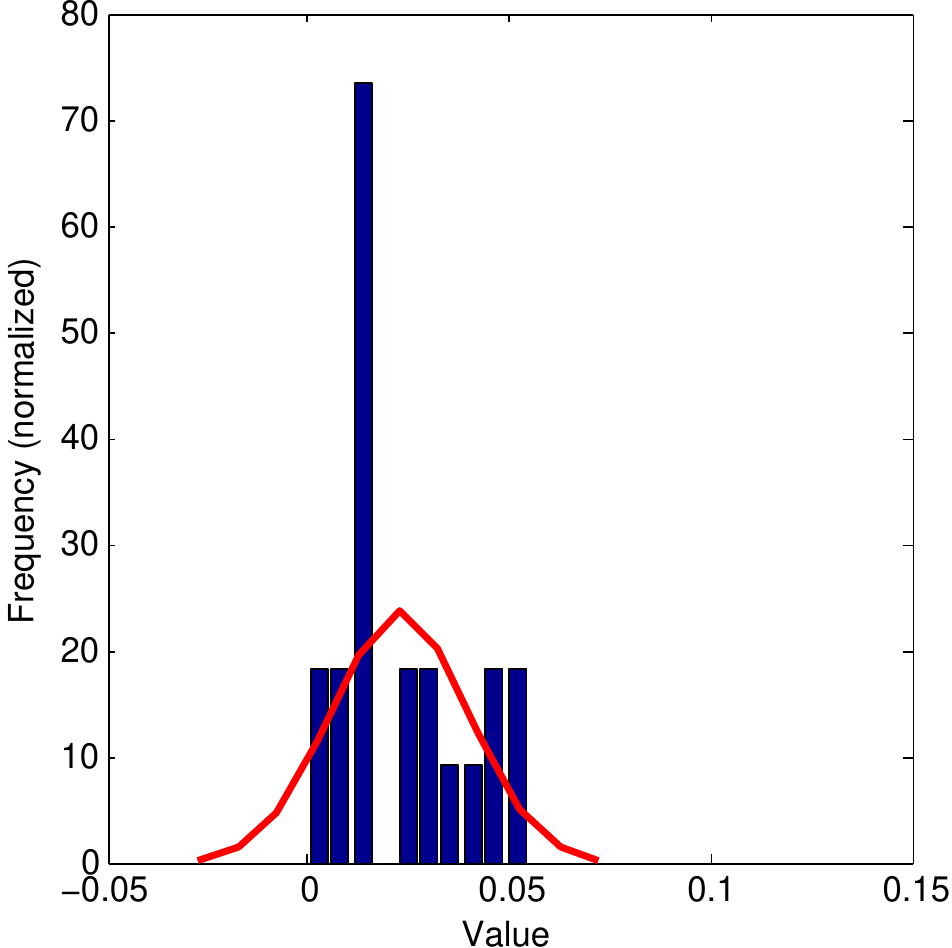}
\label{gau_where_what}}
\subfloat[]{\includegraphics[height=0.8in]{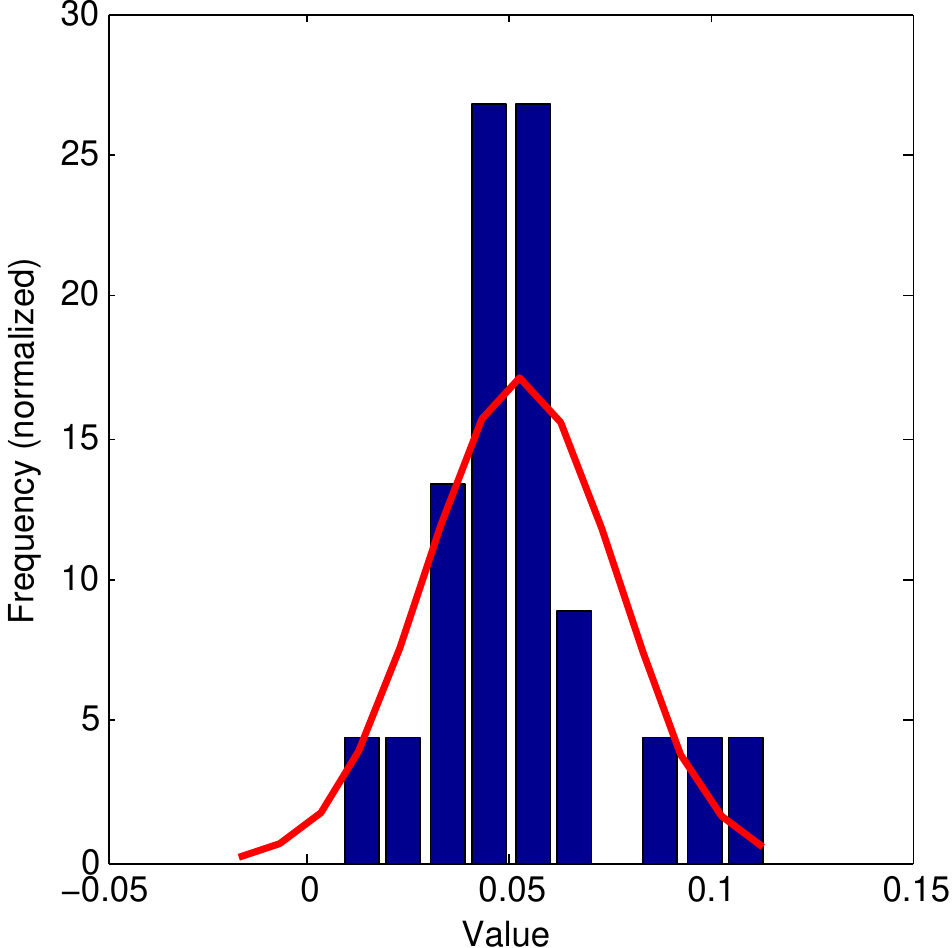}
\label{gau_text_image}}
\caption{Empirical histograms of story entity number ratios of different parts/components from one randomly selected topic (the histograms are normalized by area). Red curves show the fitted distributions. (a) Ratios of the who and where components. (b) Ratios of the who and what components. (c) Ratios of the where and what components. (d) Ratios of the text and image parts.}
\label{gaussian_dis}
\end{figure}

We assume that the ratios between the word numbers of the "who", "where" and "what" components in the text part, i.e. $\frac{M_i^{\rm{wo}}}{M_i^{\rm{wr}}}$, $\frac{M_i^{\rm{wo}}}{M_i^{\rm{wt}}}$, and $\frac{M_i^{\rm{wr}}}{M_i^{\rm{wt}}}$, follow Gaussian distributions. To verify the assumption, we do the Kolmogorov-Smirnov (KS) test for each ratio to test the goodness of how the data samples fit the Gaussian distribution. For the test, we delete the topics whose story numbers are less than 10. After this, there are 41 remaining topics. For the above three ratios, about 51.22\%, 70.73\%, and 82.93\% of the topics pass the test, and the average p-values (at the 5\% significance level) are 0.2089, 0.3761, and 0.4658 respectively. We show histograms of these three ratios for one randomly selected topic in Fig. \ref{gau_who_where}, \ref{gau_who_what} and \ref{gau_where_what} respectively.

We assume that the ratio between the total word numbers of the text part and the image part, i.e. ${\raise0.5ex\hbox{$\scriptstyle ({\sum\limits_{c_1 \in \{{\rm{wo}}, {\rm{wr}}, {\rm{wt}}\}}{M_i^{c_1}}})$}
\kern-0.1em/\kern-0.15em
\lower0.25ex\hbox{$\scriptstyle ({\sum\limits_{c_2 \in \{{\rm{face}}, {\rm{obj}}\}} M_i^{c_2}})$}}$ follows a Gaussian distribution. To verify the assumption, we do the Kolmogorov-Smirnov (KS) test on the 41 remaining large topics to test the goodness of how the data samples fit the Gaussian distribution. About 78.05\% of the topics pass the test, and the average of the p-values (at the 5\% significance level) is 0.4243. The ratio histogram of one randomly selected topic is shown in Fig. \ref{gau_text_image}.

\section{Topic Detection} \label{topic_detection}

In this section, we present our formulation of the topic detection problem, and the algorithm for optimizing a Bayesian posterior probability for the topic detection problem.

\subsection{Problem Formulation}

\begin{figure}
\centering
\includegraphics[width=3in]{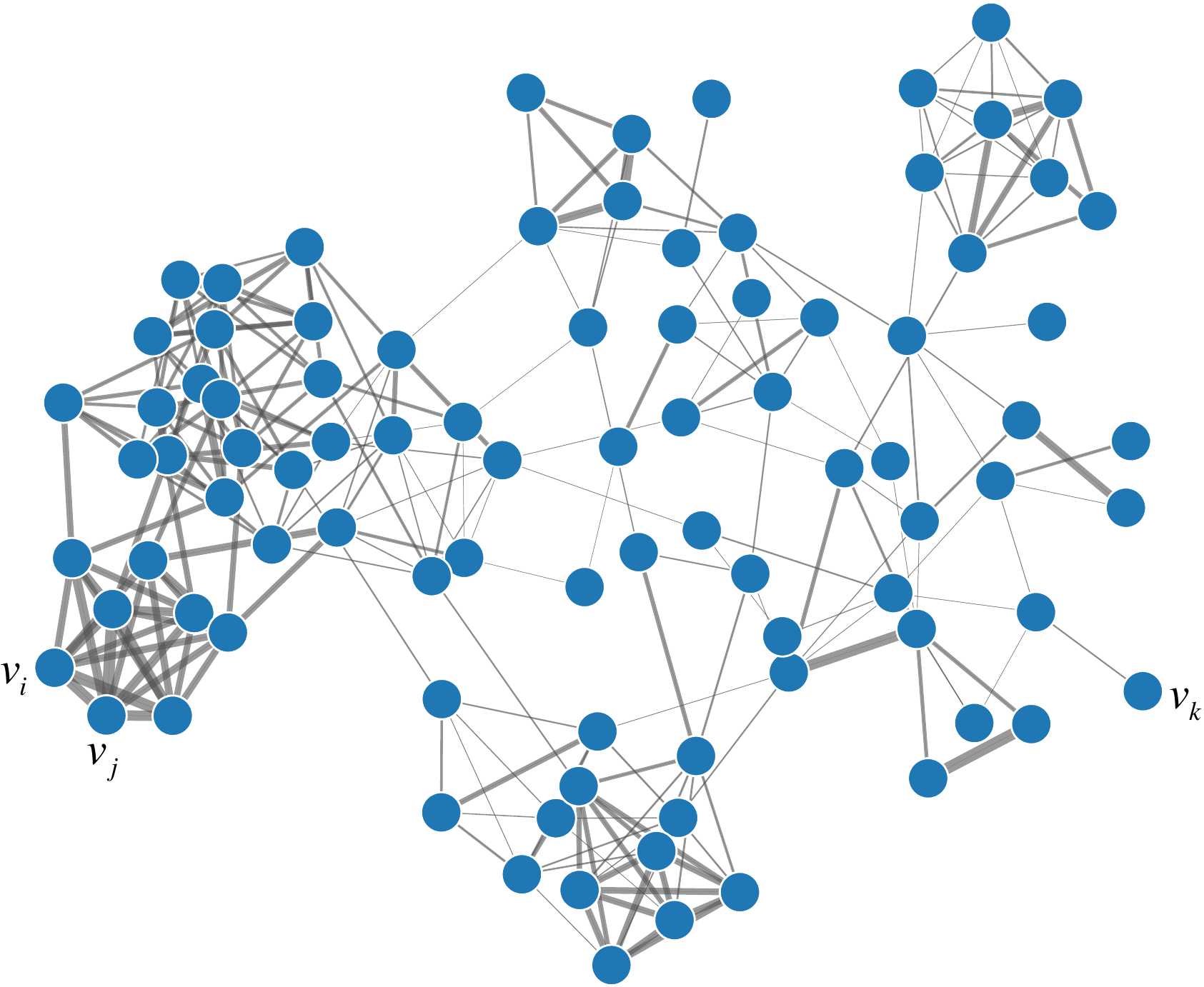}
\caption{One adjacency graph. Each vertex in the graph corresponds to one news story. The edges connect the neighboring vertices and are associated with weights corresponding to the story similarities (the edge thickness shows the story similarities). The vertices $v_i$ and $v_j$ both talk about the Oklahoma tornado topic and they are adjacent to each other in the graph. The other vertex $v_k$ which is far away from $v_i$ and $v_j$ in the graph corresponds to the story about the California High-Speed Rail project.}
\label{adj_graph}
\end{figure}

With the hierarchical AOG topic representation, our goal of topic detection is to cluster news stories that describe the same topics and obtain the AOG model parameters $\Theta$ for the topics. We pose this clustering problem as a graph partitioning problem in which news stories, as vertices in the adjacency graph, are partitioned into coherent groups. We show one example of the adjacency graph in Fig. \ref{adj_graph}. Edges in the adjacency graph are associated with certain weights corresponding to related story similarities. Partitions can be obtained by dividing the vertices into groups with specific properties and also keeping the number of edges between separated components small. Graph partitioning can help the news topic detection since even though news stories from one topic develop over time and drift the topic, they can still be grouped together through the connections between temporally adjacent stories with less changes and more similarities.

Formally, we are given a news story corpus that contains $N$ news stories, i.e. $D = \{{\bf{d}}_i; i = 1, \ldots, N\}$. The adjacency graph is defined as $\mathcal G_{\rm{ADJ}} = (V_{\rm{ADJ}}, E_{\rm{ADJ}})$ where $V_{\rm{ADJ}}$ is a set of vertices and each vertex $v_i \in V_{\rm{ADJ}}$ corresponds to one news story ${\bf{d}}_i$. $E_{\rm{ADJ}}$ is a set of edges between vertices. The clustering/partition $W$ we are trying to find given $D$ is defined as:
\begin{equation}
W = (K, \pi_K, \Theta),
\end{equation}
where $K$ is determined automatically while solving the partitioning problem and $\pi_K$ represents the $K-$partition of the adjacency graph. $\pi_K$ is defined as:
\begin{equation}
\pi_K = (V_1, ..., V_K), \bigcup\nolimits_{k = 1}^{K} {V_k} = V_{\rm{ADJ}}, V_k \cap V_j = \emptyset, \forall i \neq j.
\end{equation}
This becomes an optimization problem which can be solved by maximizing a Bayesian posterior probability:
\begin{equation}
W^* = \arg\max_{W\in\Omega}{p(W|D)} = \arg\max_{W\in\Omega}{p(D|W)p(W)}.
\end{equation}
The likelihood probability $p(D|W)$ is formulated as:
\begin{equation}
p(D|W) = \prod_{i=1}^{N}p({\bf{d}}_i; \Theta) \propto \exp\{\sum_{d_i \in D} {score}^{\rm{root}}({\bf{d}}_i; \Theta)\}.
\end{equation}
The prior probability $p(W)$ penalizes the partition number $K$ in $W$ and we formulate it as:
\begin{equation}
p(W) \propto \exp\{-\alpha N K\}.
\label{whole_num_prior}
\end{equation}
$\alpha$ is a positive parameter which acts as a threshold for grouping stories into topics. This prior helps us combine close partitions to get dense results.


\subsection{Inference by Swendsen-Wang Cuts} 

For the topic detection problem formulated above, we adopt a cluster sampling method Swendsen-Wang Cuts (SWC) \cite{swc_paper}. It is a Markov Chain Monte Carlo method which can sample the solution space $\Omega$ efficiently. An alternative method will be the expectation-maximization (EM) algorithm. But in \cite{maria_paper}, SWC is shown to be more effective than EM which finds only a local minimum.

SWC changes the labels of a group of vertices at the same time. It thus solves the coupling problem of Gibbs sampler (which flips a single vertex) by quickly jumping between local minima. SWC starts with an initial partition $\pi$, which can be the one which sets all stories to be in the same group, or can be set randomly. We denote the set of edges whose related two vertices belong to the same group under the partition $\pi$ by $E(\pi)$. The optimal clustering $W^*$ can be obtained by performing the following steps iteratively until convergence.

(1) {\emph{Determining edge status.}} Each edge $e = <v_i, v_j> \in E(\pi)$ is associated with a Bernoulli random variable $u_e\in \{0, 1\}$ which indicates the edge's on/off status and a turn-on probability $q_e$. We define:
\begin{equation}
q_e = e^{-\mathcal{D}(e)/T},
\end{equation}
where $T$ is the temperature factor and $\mathcal D(e)$ is the distance of these two vertices obtained based on the Kullback-Leibler (KL) divergence:
\begin{equation}
\mathcal{D}(e) = \sum\limits_{F \in \mathcal F}\lambda_F \cdot \dfrac{\textit{KL}(F(v_i)||F(v_j)) + \textit{KL}(F(v_j)||F(v_i))}{2},
\label{eq_edge_prob}
\end{equation}
where $F(\cdot)$ denotes one type of feature of the vertex and $\lambda_F$ is the weight for feature $F$. Here we use the distributions for the five components in the text and image parts (i.e. who, where, what, face and object) to construct the feature set $\mathcal F$. Moreover, since KL divergence is non-symmetric, we average the KL divergence of $F(v_i)$ given $F(v_j)$ and the KL divergence of $F(v_j)$ given $F(v_i)$ to get a symmetric distance measure for vertices $v_i$ and $v_j$. Based on these definitions, in this step, we set $u_e=0$ (i.e, turn $e$ off) with probability $1-q_e$ for all $e \in E(\pi)$.

(2) {\emph{Computing connected components.}} Once the states $u_e$ is determined for each edge $e \in E(\pi)$, the graph $G$ is partitioned into a set of connected components, each of which contains vertices that belong to the same group.

\begin{figure}
\centering
\subfloat[Before flipping.]{\includegraphics[height=1.5in]{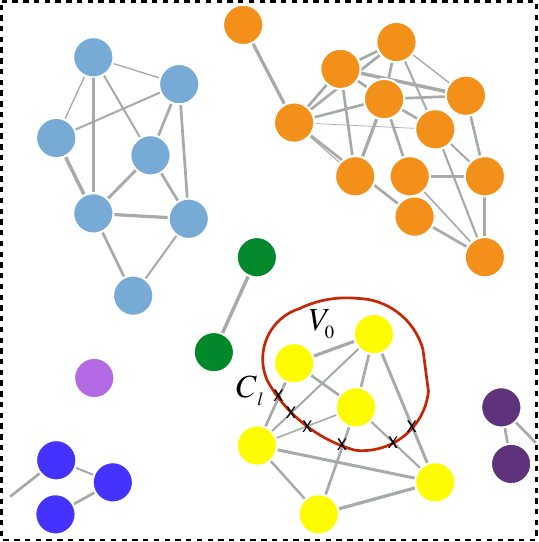}
\label{flip_before}}
\subfloat[After flipping.]{\includegraphics[height=1.5in]{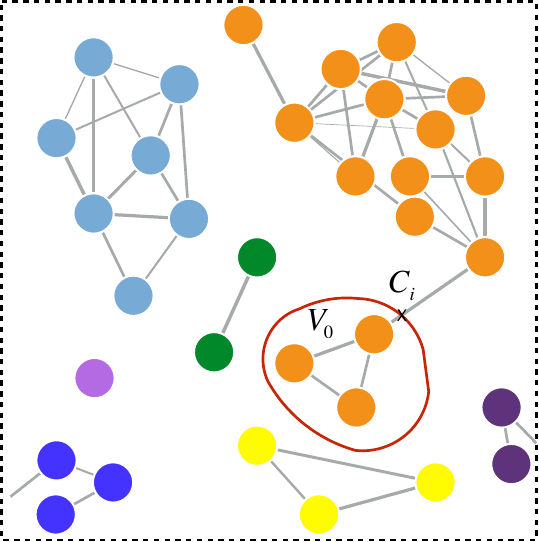}
\label{flip_after}}
\caption{SWC flips the selected component $V_0$. The cuts are marked with crosses.}
\label{swc_flip}
\end{figure}

(3) {\emph{Selecting a component and flipping it.}} Among all the connected components formed in (2), we can randomly select one component $V_0$ to flip. We show one example of $V_0$ in Fig. \ref{flip_before}. The target label for $V_0$ can be a new one that has not been used yet or just the same as any other connected components, thus allowing reversible jumps in the solution space. The current partition number is denoted as $K'$. Then the number of possible new labels for the selected component is $K'+1$. Assuming that $V_0 \subseteq V_l$ in the current partition $\pi$, we denote a series of sets
\begin{equation}
S_1 = V_1, S_2 = V_2, ..., S_l = V_l \backslash V_0, S_{K'} = V_{K'}, S_{K'+1} = \emptyset
\end{equation}
that $V_0$ can be merged with. The selected component $V_0$ can be flipped by drawing a random sample $l'$ with probability
\begin{equation}
p(l'(V_0) = i | V_0, \pi) = \dfrac{\gamma_i p(\pi_i|D)}{\sum_{j=1}^{K'+1}\gamma_jp(\pi_j|D)},
\label{equ_flip}
\end{equation}
where $\pi_i$ is the partition after assigning the label of the component $V_0$ to be $i$ and keeping other components' labels the same as in $\pi$. We also have 
\begin{equation}
\gamma_i = \prod_{e\in\mathcal{C}_i}(1-q_e),
\end{equation}
where $\mathcal C_i$ is the cuts between $V_0$ and $S_i$, i.e. $\mathcal C_i = \mathcal C(V_0, S_i) = \{<s, t>: s \in V_0, t \in S_i\}$. Two examples of the cuts are shown in Fig. \ref{swc_flip}, which are marked by the crosses. Theorem 3 in \cite{swc_paper} proved that the acceptance rate will be 1 by choosing the new label of $V_0$ by Eq. \ref{equ_flip}.

Another thing to be noted here is that when generating the adjacency graph, we can use a complete graph of $N$ vertices since each pair of news stories can be related. But this may cause problems since a complete graph of $N$ vertices has ${N \choose 2} = O(N^2)$ edges and the number of all possible solutions is exponential in the the number of edges, i.e. $O(2^{N^2})$, which requires a long convergence time. By investigating the data, however, one may observe that some story pairs have few similarities in terms of contents. Such pair of stories shall never be grouped together. Hence, graph pruning can be performed before actually running the SWC on the adjacency graph. We define a threshold $\tau$, and cut all edges $e$ whose $\mathcal{D}(e) \geq \tau$ deterministically.

Simulated annealing procedure is conducted in the optimization process. The temperature $T$ in the annealing procedure is slowly decreased according to a cooling schedule.

\section{Topic Tracking} \label{topic_tracking}

In this section, we describe our method for tracking a variable number of topics detected in certain continuous time periods. In contrast to traditional topic tracking problem \cite{TDT_book} where the topic to be tracked is provided and the task is to determine whether new-coming stories belong to the given topic, we instead link all the detected topics in different time periods to form topic trajectories over time. Then based on the trajectories, we can do further analysis such as the analysis of sentiment/emotion changes as the topic evolves.

We divide the whole news data collection into several sub-collections which consist of news stories in different time periods. Topic detection is performed within each sub-collection separately. The sub-collection set of the news corpus $D$ is denoted by $\{C_1, ..., C_M\}$ where $C_1 \cup ... \cup C_M = D$ and $M$ is the number of sub-collections. Each sub-collection contains news documents from one specific time span $t_i$. Topics extracted within each sub-collection $C_i$ are denoted by ${\bf\Theta_i} = \{\Theta_i^1, ..., \Theta_i^{K_i}\}$, where $K_i$ is the obtained topic number. 

For topic tracking, we link topics detected in the sub-collections. One optional method for solving the linking problem is to do another clustering on the detected topics using SWC. But to fast obtain the topic links, we choose to measure the similarities between topics by considering both the topic content similarities and their temporal distances, and use a threshold to decide whether they can be linked. Topic content similarity is defined based on the proposed hierarchical topic representation which models both the textual and visual information. Formally, in the tracking process, the similarity measurement to decide whether two topics can be linked is calculated as:
\begin{equation}
\begin{aligned}
Sim(\Theta_{i_1}^{k_1}, \Theta_{i_2}^{k_2}) = {\alpha_{sim}} \exp \{ - \beta_{kl}[KL({\Theta _{i_1}^{k_1}}||{\Theta _{i_2}^{k_2}}) \\ + KL({\Theta _{i_2}^{k_2}}||{\Theta _{i_1}^{k_1}})] \} + (1 - \alpha_{sim}) \exp\{-|t_{i_1} - t_{i_2}|\},
\end{aligned}
\label{eq_sim}
\end{equation}
where $i_1\neq i_2$, and $\alpha_{sim}$ and $\beta_{kl}$ are positive parameters. Note that using the proposed topic representation, each topic is composed of the image part and the text part, and they can be further divided into the "who", "where" and "what" components, and the face and object components respectively. Thus we have five components in total. Each component is represented using one model. The KL divergence of one topic given another is therefore averaged over these models:
\begin{equation}
\label{eqn_kl}
\textit{KL}(\Theta_{i_1}^{k_1}||\Theta_{i_2}^{k_2}) = \sum_{j=1}^{5} \lambda_j\textit{KL}(\Theta_{m_1}^{k_1, j}||\Theta_{i_2}^{k_2, j}),
\end{equation}
where $ \lambda_j$ is the corresponding weight for different parts. $\Theta_{i_1}^{k_1, j}$ and $\Theta_{m_2}^{k_2, j}$ are the histograms of word frequencies for the $j$-th component. After calculating the topic similarities using Eq. \ref{eq_sim}, a threshold $\tau_{link}$ can be used for pruning the links between topics to get the final topic trajectories.

\section{Experiments} \label{experiments}

\subsection{Datasets} \label{data_preparation}

Two datasets are used in our experiment:

1) \textbf{Reuters-21578.} Reuters-21578 dataset\footnote{Reuters-21578 dataset can be downloaded at \url{http://www.daviddlewis.com/resources/testcollections/reuters21578/}.} is a publicly available collection of news stories from Reuters newswire. It is widely used for the evaluation of clustering and classification methods. The dataset contains 21,758 stories which belong to 135 clusters/categories. The clusters/categories are annotated manually. Only textual information is contained in the dataset.

2) \textbf{UCLA Broadcast News Dataset.} We collected a multimedia broadcast news dataset from UCLA Library Broadcast NewsScape. Five US networks are included in the dataset: CNN, MSNBC, FOX, ABC, and CBS. It contains 379 news videos broadcasted in the time period from June 1, 2013 to June 14, 2013. The total length of the videos is about 362 hours. Several programs from each news network are included in the dataset, such as "CNN Newsroom", "CNN Situation Room", "MSNBC News Live", "FOX Morning News", "ABC Eyewitness News", "ABC Nightline", "CBS News", etc. 

\textbf{Annotation: } We annotate the UCLA Broadcast News Dataset for topic detection and tracking. One annotation choice can be letting the annotators manually group the new stories based on their related topics \cite{1621449, Allan98topicdetection}. However, this will be a hard task and the results may not be accurate since there can be hundreds of news topics even in one week and the annotators can hardly remember all the previously found topics during annotation. So instead of this, we choose to build the ground-truth by letting annotators decide whether a pair of stories belong to the same topic or not. The topic granularity is chosen to be at the event level, like the definition in the TDT system \cite{Allan98topicdetection}. In other words, two stories talking about the same event (or closely related ones) belong to the same topic. Since it takes a long time to annotate all story pairs (about $N^2$ pairs for $N$ stories), we choose to annotate a subset selected from the whole story pair collection. We first compute the cosine distances between the two stories in the same pair, and then select 10,000 story pairs to be annotated randomly from the pair set where all pair distances are within the range $[0.6, 0.9]$. This specific range is chosen for the reason that the corresponding story relations are ambiguous compared to other ranges. Three annotators are involved in the annotation and for each story pair we treat the relation that most annotators agree as the ground-truth relation.

This dataset is mainly used for quantitative evaluation of our method. To show how our method work on large-scale datasets qualitatively, we also apply our method to more news data from the UCLA Library Broadcast NewsScape.

\subsection{Implementation Details} \label{preprocessing_new}

The implementation details (including the preprocessing procedures) for the two datasets used in the experiment are described below:

1) \textbf{Reuters-21578:} In the experiment, stories with multiple cluster labels are discarded and for the remaining stories, only those from the largest 10 clusters are selected \cite{Xie}.

2) \textbf{UCLA Broadcast News Dataset:} We utilize texts from both video frames and closed captions (CC). Text extraction on video frames is performed using optical character recognition (OCR) based on Google OCR engine Tesseract \cite{Tesseract}, and the results are further refined using the spatial-temporal relations between frames. News CC consist of several stories in one single continuous text stream. Story segmentation needs to be performed to divide the CC into stories. In CC, some special markers are used as the indicators of story boundaries, such as "$>>>$". Moreover, many news programs insert commercials between stories with special formats of letter cases and indentations. Thus we also do commercial detection based on these special formats. Using the special boundary markers and commercial detection results, most stories boundaries are determined. For the remaining boundaries, we train a classifier using Support Vector Machine to decide the boundary locations based on features including the boundary key words (such as "coming up", "still ahead"), and similarities of sentences near the boundaries.

For the news videos, we extract the keyframes by removing the commercial frames, redundant frames and anchor frames. Commercial frames can be specified using the aforementioned commercial detection results from CC. Redundant frames are those perceived to be similar to the previous frames. We use the frame histograms to decide whether one incoming frame is similar to the previously detected non-redundant frame. After removing the commercial and redundant frames, we further detect anchor frames among the remaining frames. Anchor frames are those containing the news anchors. They usually appear repeatedly in the video. We detect the anchor frames by exploiting features from two aspects: anchor frames' backgrounds (they usually show the news studios and thus are similar in the videos), and anchors' faces. Similar backgrounds and faces are grouped by clustering. We can then check the clusters' time distribution and decide whether the corresponding frames are anchor frames or not.

\begin{figure}[]
\centering
\includegraphics[width=3in]{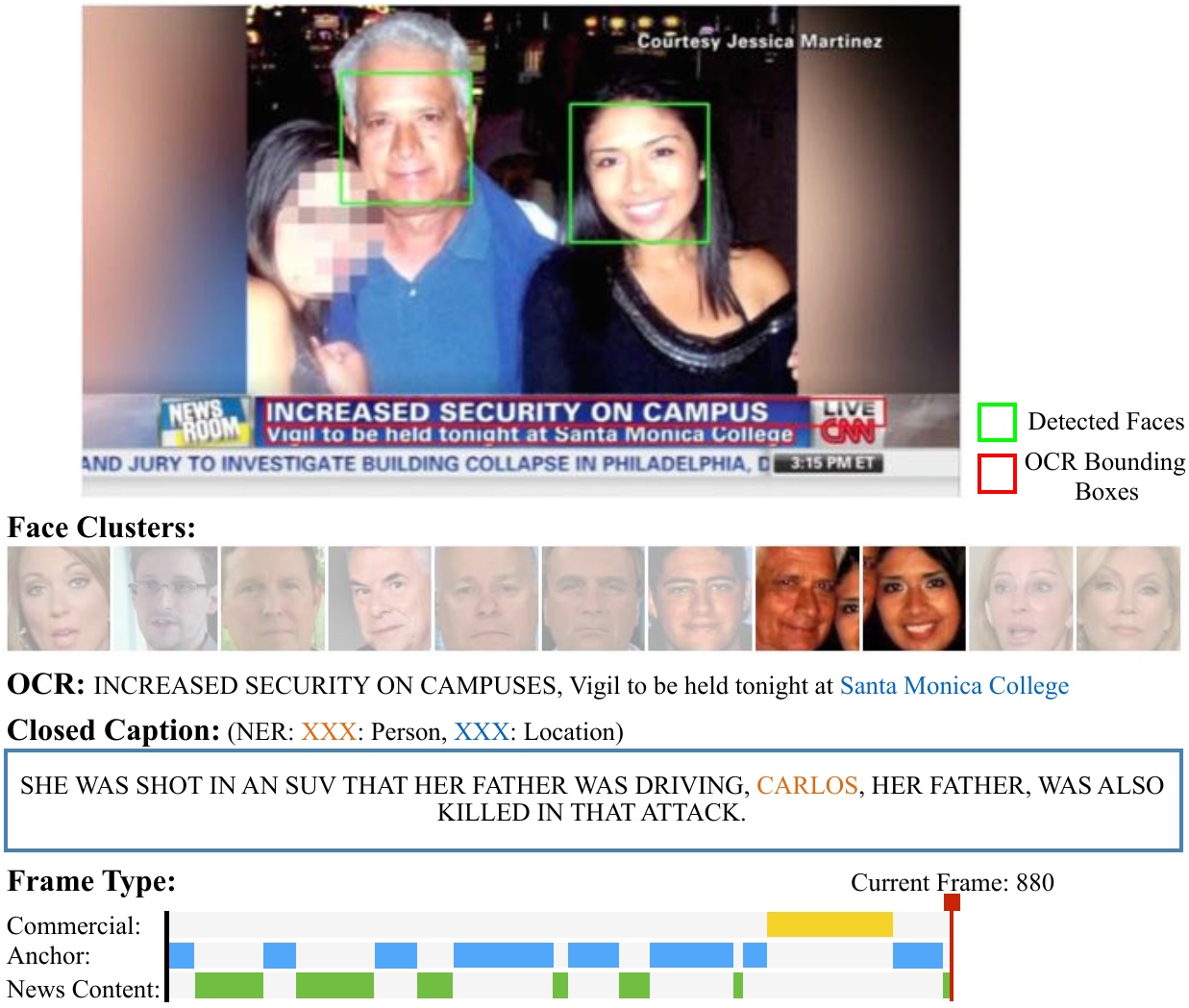}
\caption{An example showing the preprocessing results, including those for OCR, NER, face detection and clustering, commercial detection (marked by "Commercial" label in the "Frame Type" part), and anchor frame detection (marked by the "Anchor" label). The horizontal axis for the "Frame Type" part is the time axis.}
\label{data_processing}
\end{figure}

Fig. \ref{data_processing} illustrates the results that can be obtained in the previous preprocessing procedure. After preprocessing, we got 3,633 news stories including 577,721 words and 36,810 keyframes. The whole collection contains 24,036 unique word terms. 

\subsection{Experiment I: Topic Detection}

In this experiment, we conduct topic detection experiments on both the Reuters-21578 dataset and the UCLA Broadcast News Dataset.

\subsubsection{Results on Reuters-21578} \label{reuters_comparison}

We compare the proposed topic detection method with other story/document clustering methods on the news dataset Reuters-21578. Only texts are used in the comparison.

\textbf{Evaluation Protocol.} On Reuters-21578, we follow the evaluation protocol in \cite{Xie, dengcai}. Two metrics are used to evaluate the clustering performance, i.e. accuracy, and normalized mutual information. To compute the accuracy, the obtained clusters are mapped to the ground-truth clusters in the dataset. The accuracy is then defined as the percentage of the documents that have the correct cluster labels after mapping. The mutual information measures the mutual dependence of the ground-truth cluster assignments and the obtained clustering assignments for the documents. The normalized mutual information is a normalized version of this measure. More details of the definitions of these two metrics can be found in \cite{dengcai}.

\textbf{Baseline Methods.} Several baseline methods are included in the comparison, namely:
\begin{itemize}
\item K-means and Normalized Cuts (NC) \cite{Shi:2000:NCI:351581.351611}, which are widely used clustering and graph partitioning algorithms.
\item Nonnegative-Matrix-Factorization (NMF) based clustering \cite{Xu:2003:DCB:860435.860485}, Latent Semantic Indexing (LSI) \cite{Hofmann:1999:PLS:312624.312649}, and Locally Consistent Concept Factorization (LCCF) \cite{dengcai}, which are factorization based methods that are very effective in document clustering.
\item LDA related methods: 1) LDA + K-means \cite{Xie}: using LDA to learn the topics and the topic distribution for each document, and then clustering using K-Means based on these distributions; 2) LDA + Naive \cite{Xie}: using LDA to learn the topics and the documents' topic distributions, and then treating the label of the most dominant topic as the cluster label for each document.
\item Multi-grain clustering topic model (MGCTM) \cite{Xie} which integrates document clustering and topic modeling. It has the best clustering result on Reuters-21578 so far.
\end{itemize}
The inputs of these methods in the comparison are the documents' tf-idf vectors \cite{Xie, dengcai}. Stop words are removed from the documents. These methods all require the cluster number to be specified in the input. Thus for these methods, we set the cluster number $K = 10$ in the experiment, which equals the ground-truth cluster number in the dataset. Please refer to \cite{Xie} for other detailed settings of these algorithms.

\textbf{Parameter Settings of Our Method.} To compare with the baseline algorithms, in our method, we add a Gaussian prior term with the mean $\mu = 10$ and variance ${\sigma ^2} = 0.5$ to Eq. \ref{whole_num_prior} to make the sampling process converge to the state where the cluster number equals $10$. The parameter $\alpha$ in Eq. \ref{whole_num_prior} is set as $\alpha = 0.2$. The weights $\{\lambda_F, F \in \mathcal F \}$ in Eq. \ref{eq_edge_prob} are set as: $\lambda_{F_{who}} = 0.1$, $\lambda_{F_{where}} = 0.1$, $\lambda_{F_{what}} = 0.4$, $\lambda_{F_{face}} = 0.1$ and $\lambda_{F_{object}} = 0.3$. The threshold $\tau$ used for graph pruning is set as $\tau = 160$.

\begin{table}
\caption{Clustering Performance of different methods on Reuters-21578.}
\begin{center}
\begin{tabular}{|c|c|c|c|}
\hline
					& Clustering 			& Normalized Mutual \\
					& Accuracy(\%)			& Information(\%) \\
\hline
K-Means 				& 35.02 							& 35.76 \\
\hline
NC \cite{Shi:2000:NCI:351581.351611} 					& 26.22								& 27.40 \\
\hline
NMF \cite{Xu:2003:DCB:860435.860485}				& 49.85 							& 35.89 \\
\hline
LSI \cite{Hofmann:1999:PLS:312624.312649}				& 42.00								& 37.14 \\
\hline
LCCF 	\cite{dengcai}			& 33.07 							& 30.45 \\
\hline
LDA + K-means \cite{Xie} 	& 29.73 							& 36.00 \\
\hline
LDA + Naive \cite{Xie}		& 54.88 							& 48.00 \\
\hline
MGCTM \cite{Xie}			& 56.01 							& 50.10 \\
\hline
our method 		& \textbf{67.19}						& \textbf{51.97}	 \\
\hline
\end{tabular}
\end{center}
\label{clustering_performance_reuters}
\end{table}

\textbf{Comparison Results.} Table \ref{clustering_performance_reuters} shows the comparison results of different methods on Reuters-21578. It can be seen from the results that our approach is better than the other methods in terms of both the clustering accuracy and the normalized mutual information. This is because our method uses the AOG representation which organizes the information in a hierarchical way and embeds the contexts between different components. The cluster sampling method SWC also plays an important role in getting the optimal solution. Other methods generally use the basic word distributions and most of the solutions they get are not global optimal.

\subsubsection{Results on UCLA Broadcast News Dataset}

\begin{figure}
\centering
\includegraphics[width=3.3in]{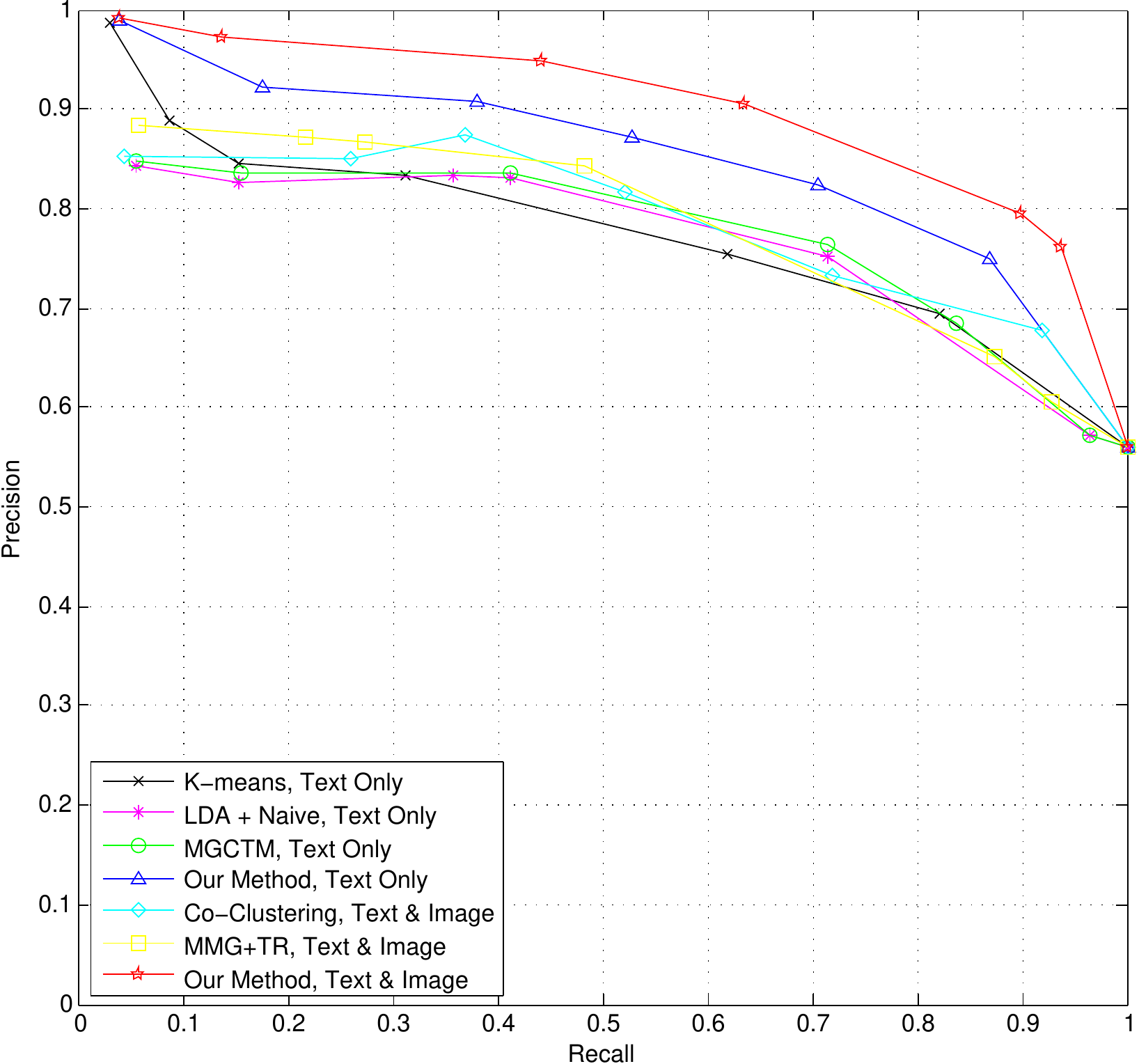} 
\caption{Precision-recall curves of different methods for topic detection on UCLA Broadcast News Dataset.}
\label{roc}
\end{figure}

We conduct both qualitative and quantitative evaluations of our topic detection method on the UCLA Broadcast News Dataset. In the qualitative experiment, we show the topics detected by our method. In the quantitative experiment, we compare our method's topic detection performance with other algorithms.

\textit{\textbf{1) Qualitative Evaluation.}} We conduct the joint image-text topic detection experiment on the whole dataset, i.e. clustering using both textual and visual information from all news stories in the dataset.

\textbf{Parameter Settings.} The parameter $\alpha$ in Eq. \ref{whole_num_prior} is set as $\alpha = 10$. The cluster numbers for grouping the faces and object patches in Section \ref{topic_representation} are set as $1,000$ and $1,500$ respectively. We also delete clusters with a small number of patches. The remaining cluster numbers for face and object are $708$ and $1,316$ respectively. Other parameter settings are the same as those in Section \ref{reuters_comparison}.

\begin{figure*}
\centering
\includegraphics[width=6.8in]{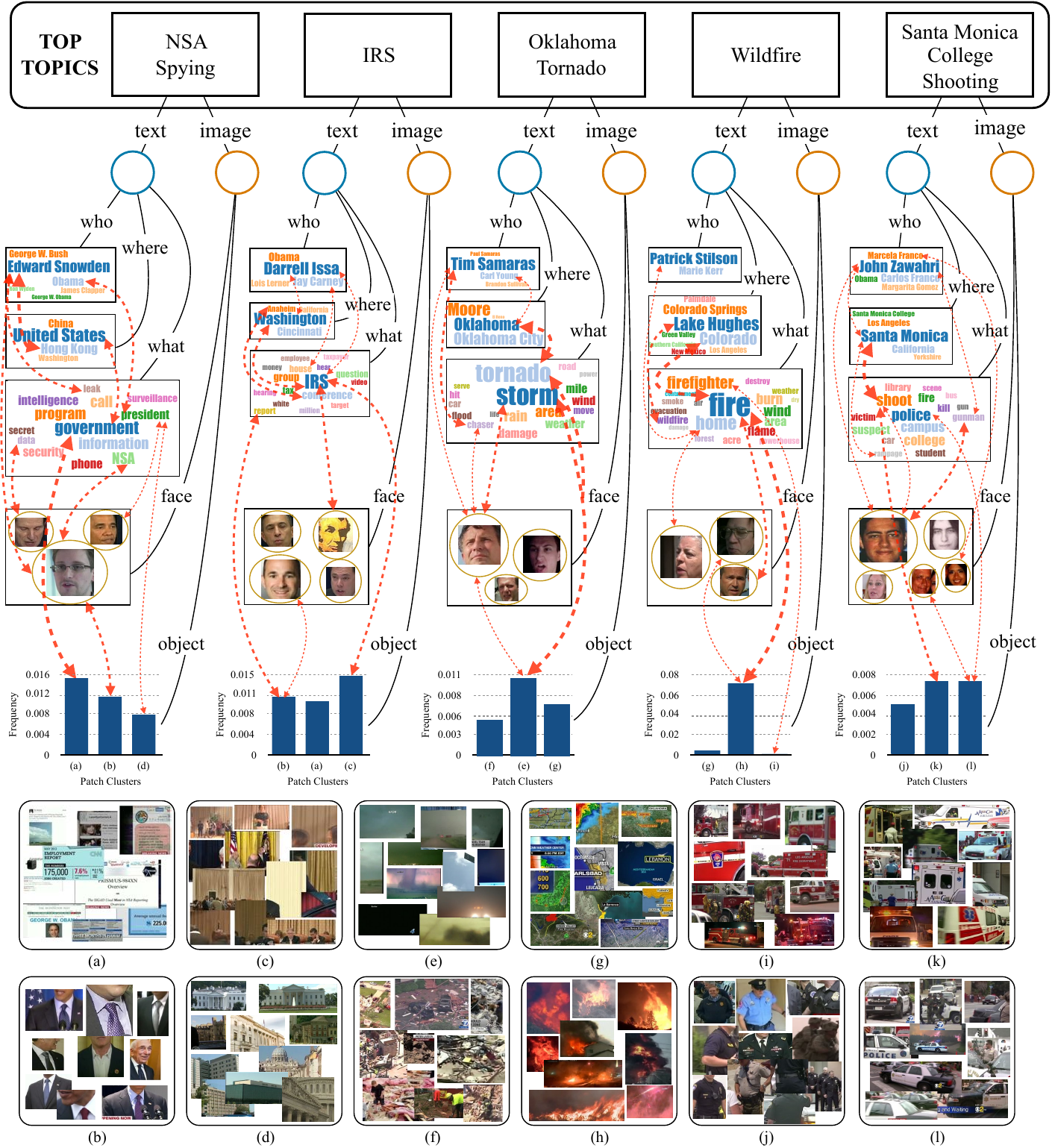}
\caption{Top five topics detected in the dataset we collected. Each topic is composed of the text part and image part. The text part is further divided into three subcomponents, i.e. "who", "where" and "what", and the top words for each component are shown with their sizes proportional to the frequencies. The image part is further divided into two subcomponents, i.e. face and object. The top faces and objects are also shown. Each face patch along with the circle outside it represent one face cluster. Their sizes are proportional to the frequency in the corresponding topic. Each object cluster is shown in one square with several object patches at the bottom part of the figure. The objects' frequencies in the topic are shown by the curves above the squares. The dashed red lines show the top co-occurring pairs between different components and the thickness of each line is proportional to the related pair frequency.}
\label{top_topic}
\end{figure*}

\textbf{Topic Detection Results.} We show the detected top five topics in Fig. \ref{top_topic}. Topic 1 talks about the news that Edward Snowden leaked information from National Security Agency (NSA). Topic 2 is about the IRS scandal, including the discussion on the misuse of taxpayers' money and the related hearing. Topic 3 mainly talks about the Oklahoma tornado, including its development, the damage it caused, and the storm chasers' stories. Topic 4 is about the wildfires, which also includes the fire development and the related damages. Topic 5 is about the Santa Monica College shooting rampage, and the related gunman and victims' stories are also included. We can see from the figure that the obtained structured results can clearly describe the related topics. The involved people's names and face patches, the related locations, the key objects, the descriptions about the event, as well as the co-occurrence relations between them (represented by the dashed red lines) are all shown in the structure.

\textit{\textbf{2) Quantitative Evaluation.}} We also conduct quantitative evaluation on the proposed joint image-text topic detection method.

\textbf{Evaluation Protocol.} Using the annotated story pairs, we draw precision-recall curves for different topic detection methods in the evaluation. The precision is calculated as the fraction of story pairs that actually belong to one topic out of those that are computed to be. The recall is the fraction of story pairs that are computed to belong to one topic out of those that actually do.

\textbf{Baseline Methods.} Among the baseline methods used in \ref{reuters_comparison}, we select some methods with better performance, including LDA + Naive and MGCTM. We also include the widely used k-means algorithm. These algorithms are all single-modal, so their inputs in the experiment are the stories' textual information, i.e. the stories' tf-idf vectors. Multi-modal baseline methods are also included in the comparison, including the multimodal co-clustering method in \cite{1621449}, and the multi-modality graph with topic recovery method (MMG+TR) in \cite{6877623}. For these method, we set a sequence of cluster numbers in the experiment to generate the precision-recall curves.

\textbf{Parameter Settings of our method.} To generate the precision-recall curve, we change the parameter $\alpha_K$ in Eq. \ref{whole_num_prior} for our method. Other parameter settings are the same as those in the qualitative experiment. To compare with the single-modal methods, we also conduct experiments where only the text information is included.

\textbf{Comparison Results.} Fig. \ref{roc} shows the precision-recall curves for different methods. As we can see from the figure, based on merely text information, our method has better performance than the other single-modal methods. This shows that the proposed hierarchical AOG topic representation and the clustering sampling method we use can help generate better topics. With the visual information added, our performance gets further improved, showing the effectiveness of our method which jointly models the text and visual information.

%

\begin{figure*}
\centering
\includegraphics[width=6in]{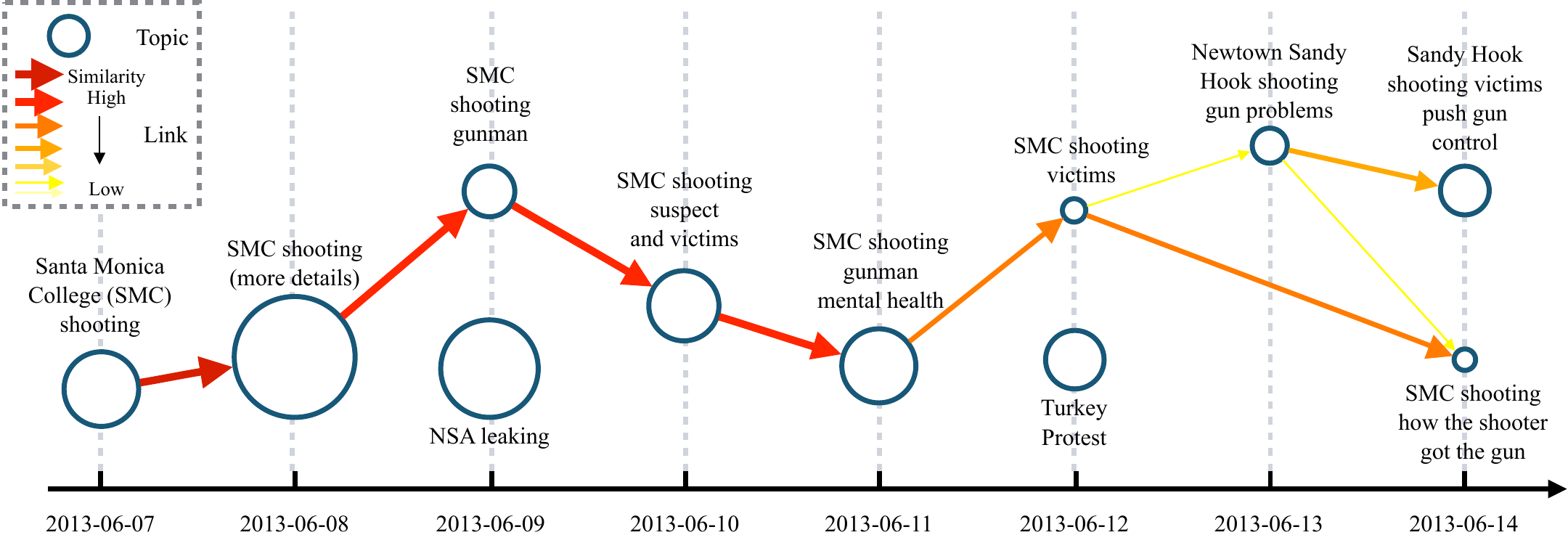}
\caption{Topic tracking result of the event Santa Monica Shooting. Each circle represents one topic and the circle size is proportional to the size of the topic, i.e. the volume of corresponding news stories. Thicker links represent greater similarities between topics.}
\label{tracking_shooting}
\end{figure*}

\subsection{Experiment II: Topic Tracking}

In this experiment, we conduct topic tracking experiments on the UCLA Broadcast News Dataset. Both qualitative and quantitative evaluations of our method are included in the experiment.

\textit{\textbf{1) Qualitative Evaluation.}} To show that our topic tracking method can generate meaningful topic trajectories, we conduct the qualitative evaluation experiment.

\textbf{Parameter Settings.} To track topics over time, we divide the whole collection of news stories in the UCLA Broadcast News Dataset into 14 sub-collections each of which contains news stories from one day. Topic detection is firstly performed within each sub-collection. Then given the detected topics, we further do topic tracking, which links topics over time and generate topic trajectories. The parameter $\alpha_{sim}$ and $\beta_{kl}$ in Eq. \ref{eq_sim} are set as $\alpha_{sim} = 0.8$ and $\beta_{kl} = 0.005$ respectively. The weights $\{\lambda_i; i=1, ..., 5\}$ in Eq. \ref{eqn_kl} are set as $\{0.1, 0.1, 0.4, 0.1, 0.3\}$. The threshold $\tau_{link}$ for selecting links between topics is set as $\tau_{link} = 0.7$.

\begin{figure}[!htb]
\centering
\subfloat[The who part.]{\includegraphics[width=3in]{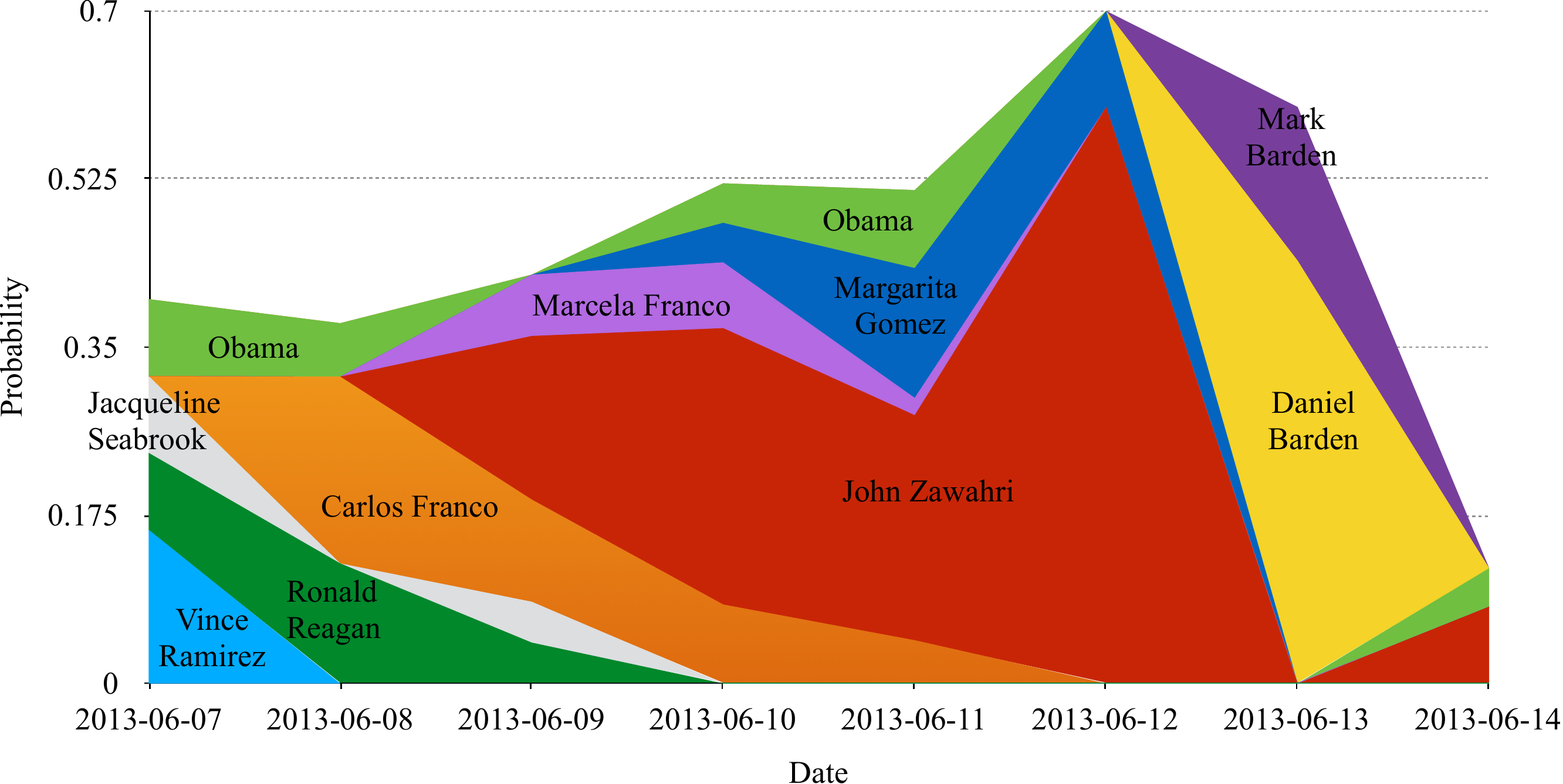}
\label{shooting_people}}\\
\subfloat[The where part.]{\includegraphics[width=3in]{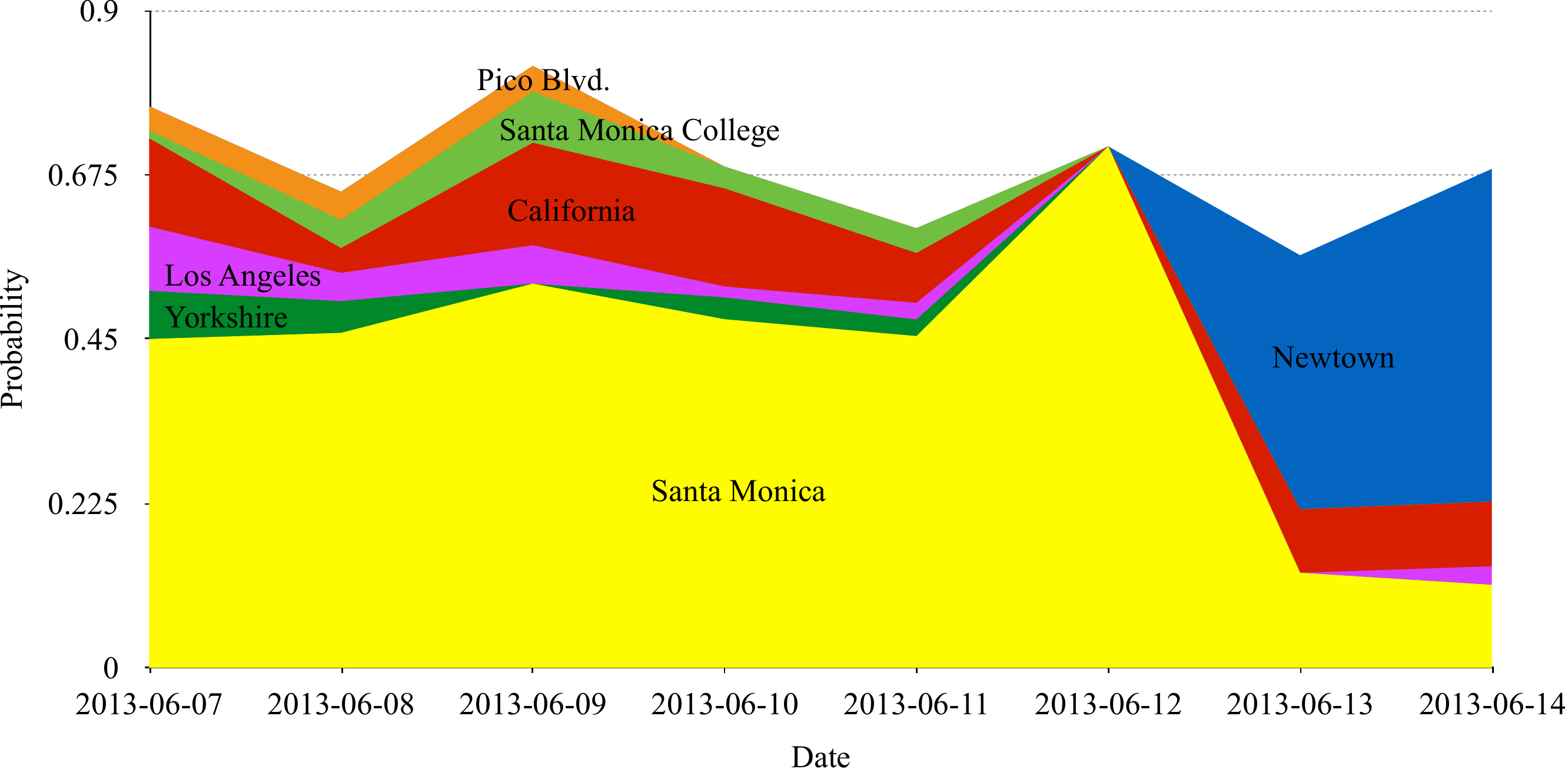}
\label{shooting_location}}\\
\subfloat[The what part.]{\includegraphics[width=3in]{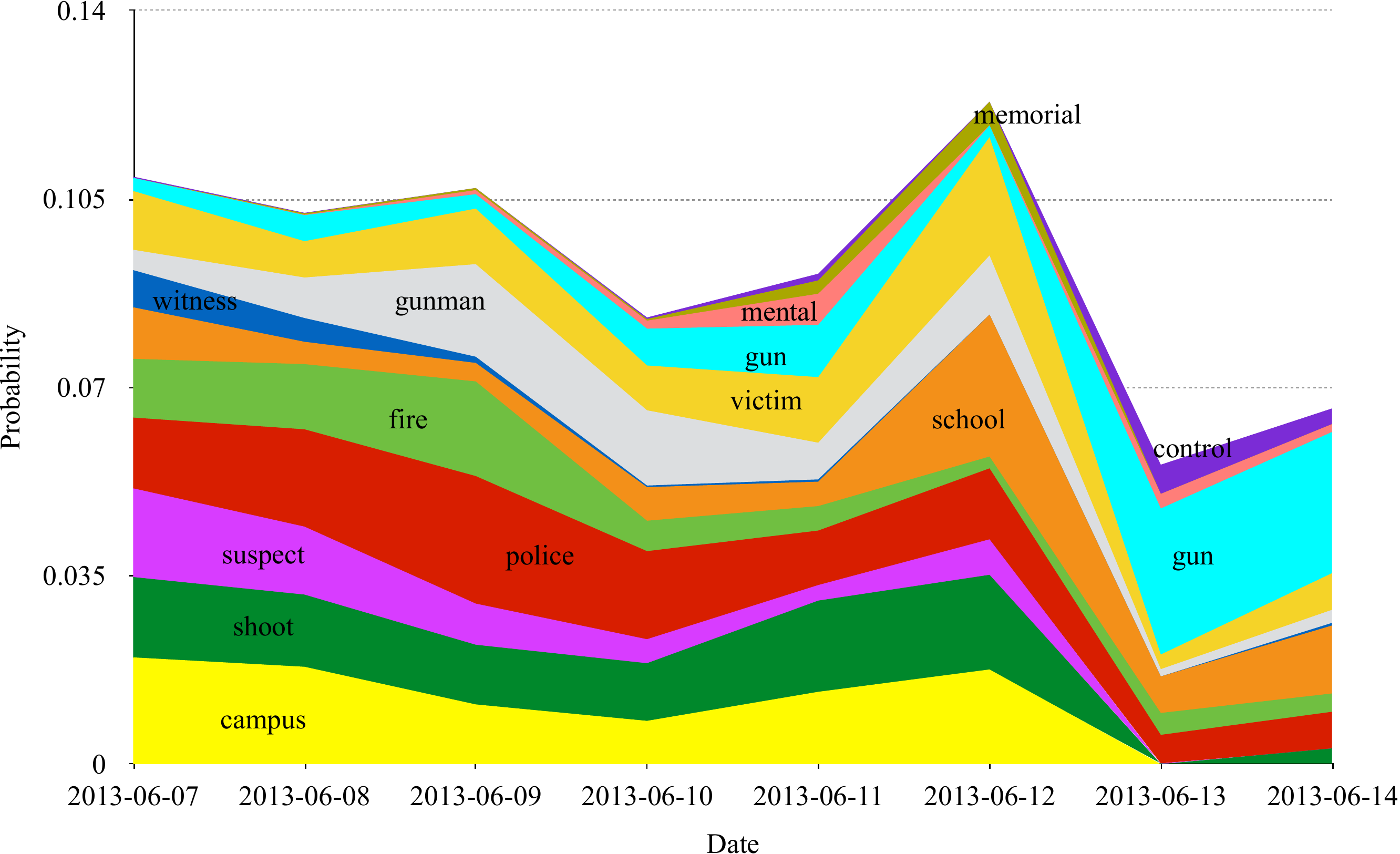}
\label{shooting_event}}
\caption{The text part of the topics corresponding to the trajectory shown in Fig. \ref{tracking_shooting}. Top words' probabilities along the time span are shown.}
\label{tracking_shooting_topics}
\end{figure}

\begin{figure}[!htb]
\centering
\subfloat[The face part.]{\includegraphics[width=3.2in]{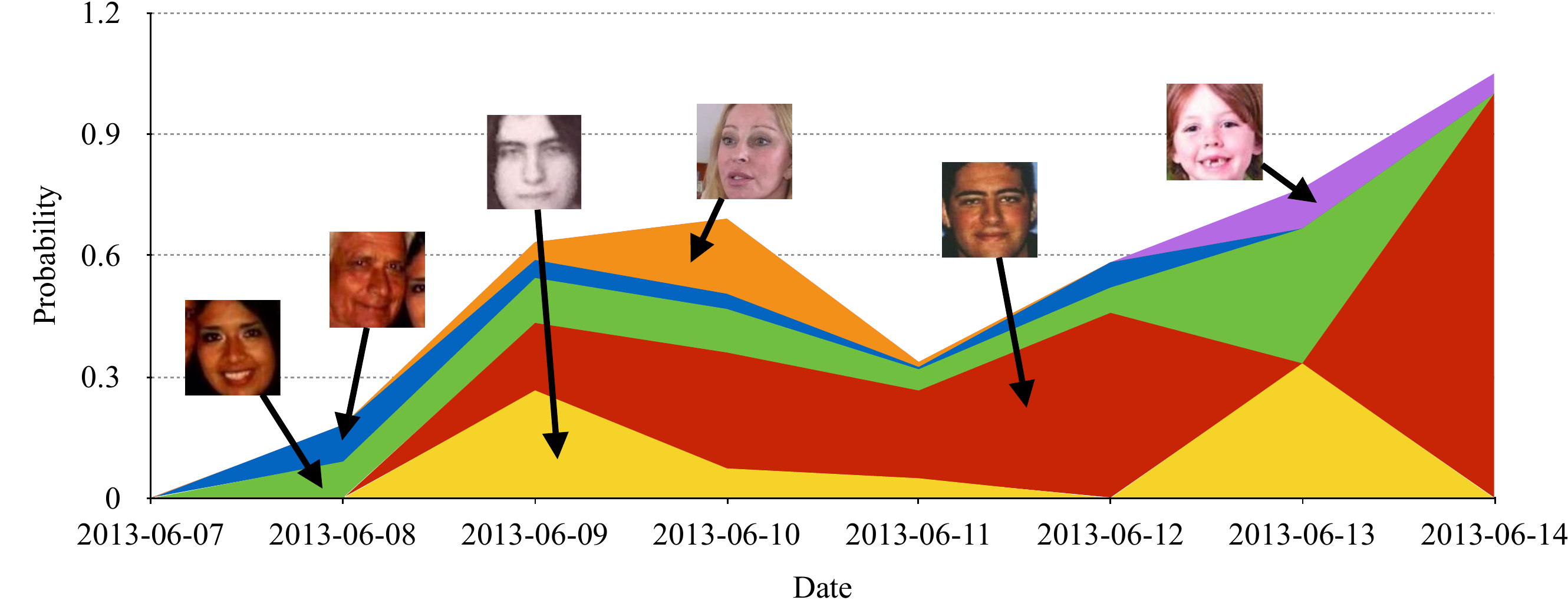}
\label{shooting_face}}\\
\subfloat[The object part.]{\includegraphics[width=3.2in]{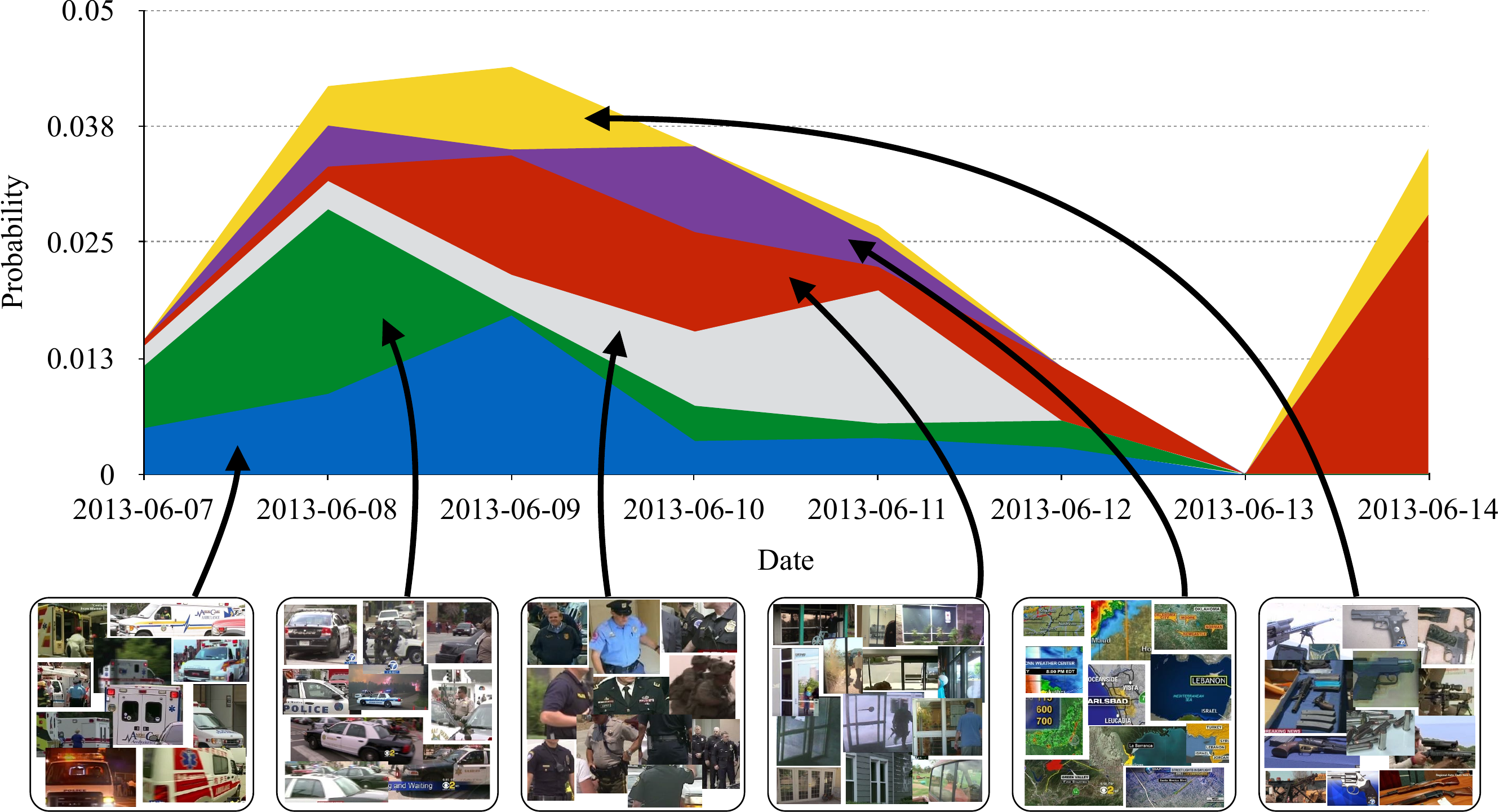}
\label{shooting_object}}
\caption{The image part of the topics corresponding to the trajectory shown in Fig. \ref{tracking_shooting}. Top faces'/objects' probabilities along the time span are shown.}
\label{tracking_shooting_topics_image}
\end{figure}

\textbf{Topic Tracking Results.} One topic tracking trajectory about the Santa Monica College shooting is shown in Fig. \ref{tracking_shooting}. In the figure each circle corresponds to one detected topic. The topics are summarized in several words here for space constraints. The size of the circle is proportional to the topic size. For the links between topics, thicker ones means greater similarities between topics. The descriptions of the text part and the image part for the corresponding topics in the trajectory are shown in Fig. \ref{tracking_shooting_topics} and Fig. \ref{tracking_shooting_topics_image} respectively. The probabilities of the top textual and visual words over time are shown in the figure.

\begin{figure}[!htb]
\centering
\includegraphics[width=3in]{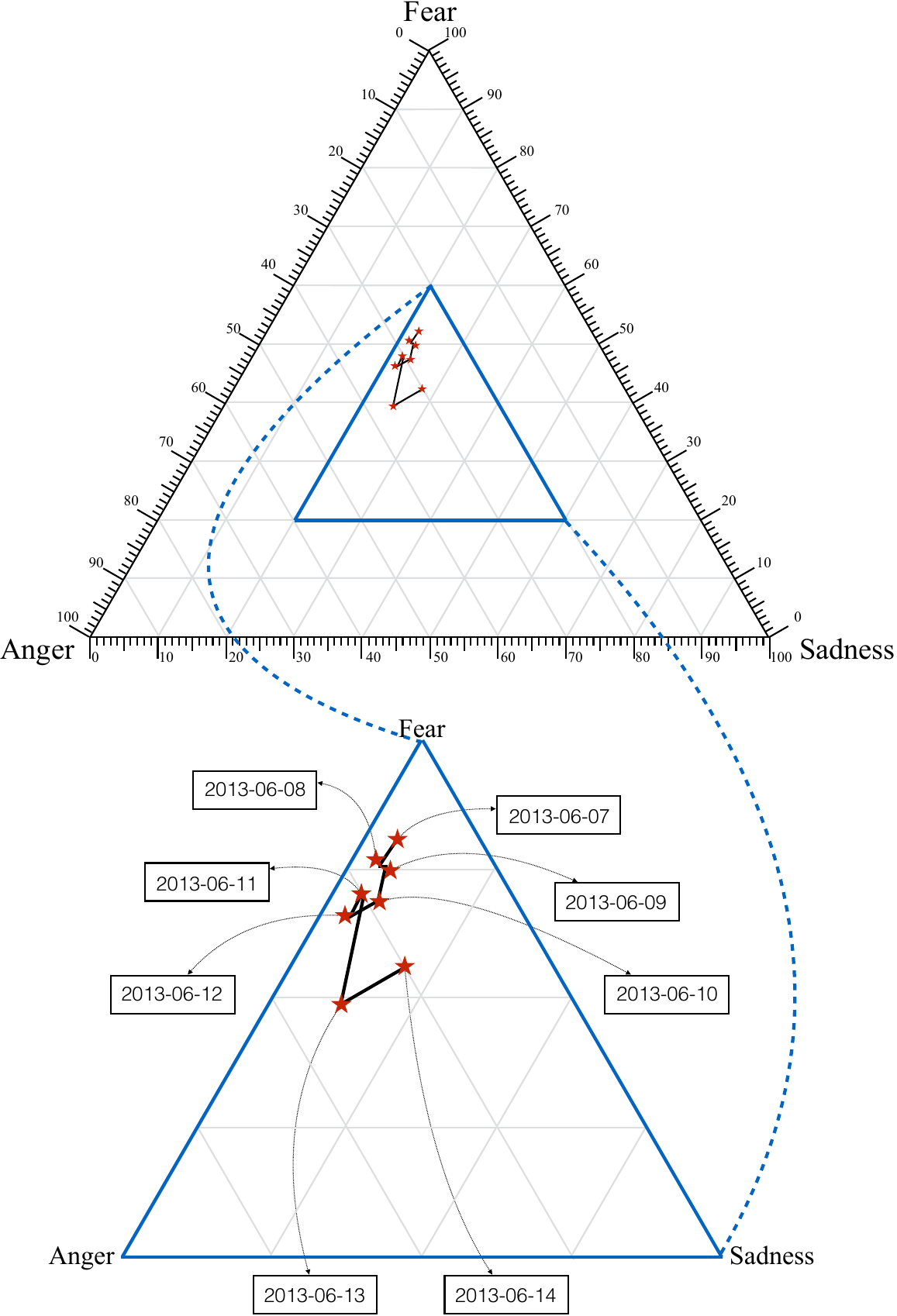}
\caption{Emotion analysis on the Santa Monica Shooting. The ternary plot on three emotional variables (fear, anger and sadness) shows the emotional changes in the news stories as the topic goes on.}
\label{shooting_emotion}
\end{figure}

Based on the tracking result, we also analyze the emotional changes as the topic developed. The NRC Emotion Lexicon \cite{Mohammad13} is used for the emotion analysis. Three emotional variables, i.e. fear, anger and sadness, are included in the analysis and the ternary plot on these variables is shown in Fig. \ref{shooting_emotion}. From these figures we can see that at the beginning, when the shooting happened, news stories mainly describe the shooting scenario and expressed people's fear mostly. Later when the suspect was found, more anger is shown in the news stories. When victims' stories were told later, sadness became dominant. From these results, we can clearly see how news media reported the event and what emotions they want to express. They also show that our tracking method can generate meaningful tracking trajectories.

\textit{\textbf{2) Quantitative Evaluation.}} We also conduct quantitative evaluation on the proposed multimodal topic tracking method.

\textbf{Evaluation Protocol.} For topic tracking, we also use the precision-recall curves to compare different methods. The annotated story pairs are used as the ground-truth data.

\textbf{Baseline Methods.} We include two baseline methods in the comparison, namely:
\begin{itemize}
\item Dynamic topic model (DTM) \cite{dynamic_topic_model} which models topic changes over time.
\item Topic chain method \cite{Kim:2011:TCU:1964750.1964765} which generates topics in different time periods using LDA and links these topics to form topic chains.
\end{itemize}
These two methods are both single-modal. For DTM, we set different topic numbers in the experiment to generate its precision-recall curve. For the topic chain method, we set the topic number in each time period as $50$ and use a sequence of similarity threshold when building the topic chains.

\textbf{Parameter Settings of our method.} To generate the precision-recall curve, we also change the parameter $\tau_{link}$ in the experiment.

\begin{figure}[!htb]
\centering
\includegraphics[width=3.3in]{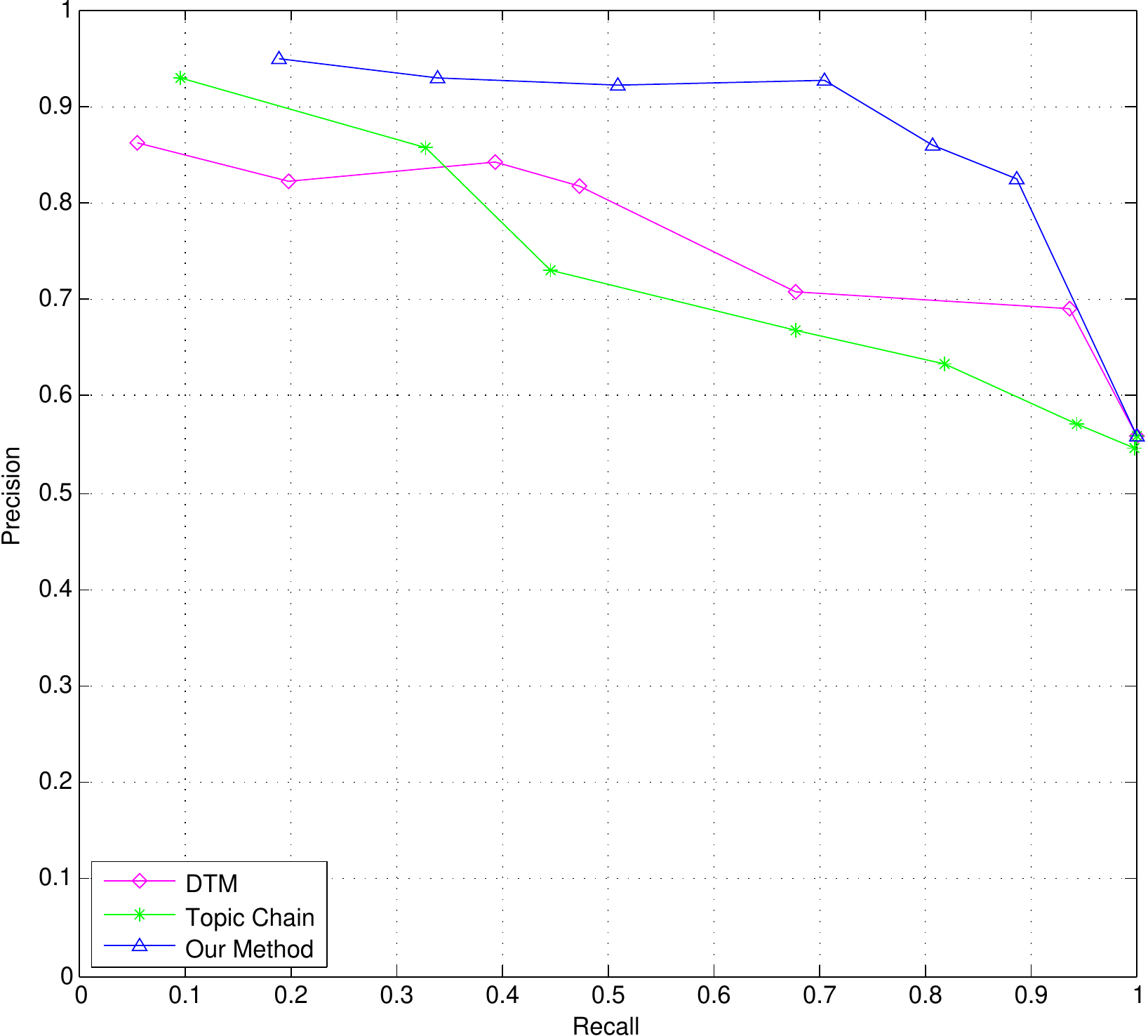} 
\caption{Precision-recall curve of the proposed topic tracking method and comparisons with other methods on UCLA Broadcast News Dataset.}
\label{roc_tracking}
\end{figure}

\textbf{Comparison Results.} Fig. \ref{roc_tracking} shows the precision-recall curves for our tracking method and the two baseline methods. Our method outperforms the baseline methods since both textual and visual information are included in the tracking process. Moreover, our topic detection method can generate meaningful topics, which is also an important factor for the topic tracking performance.

\subsection{Experiment III: Large-Scale Topic Detection and Tracking}

\begin{figure*}[!htb]
\centering
\includegraphics[width=7in]{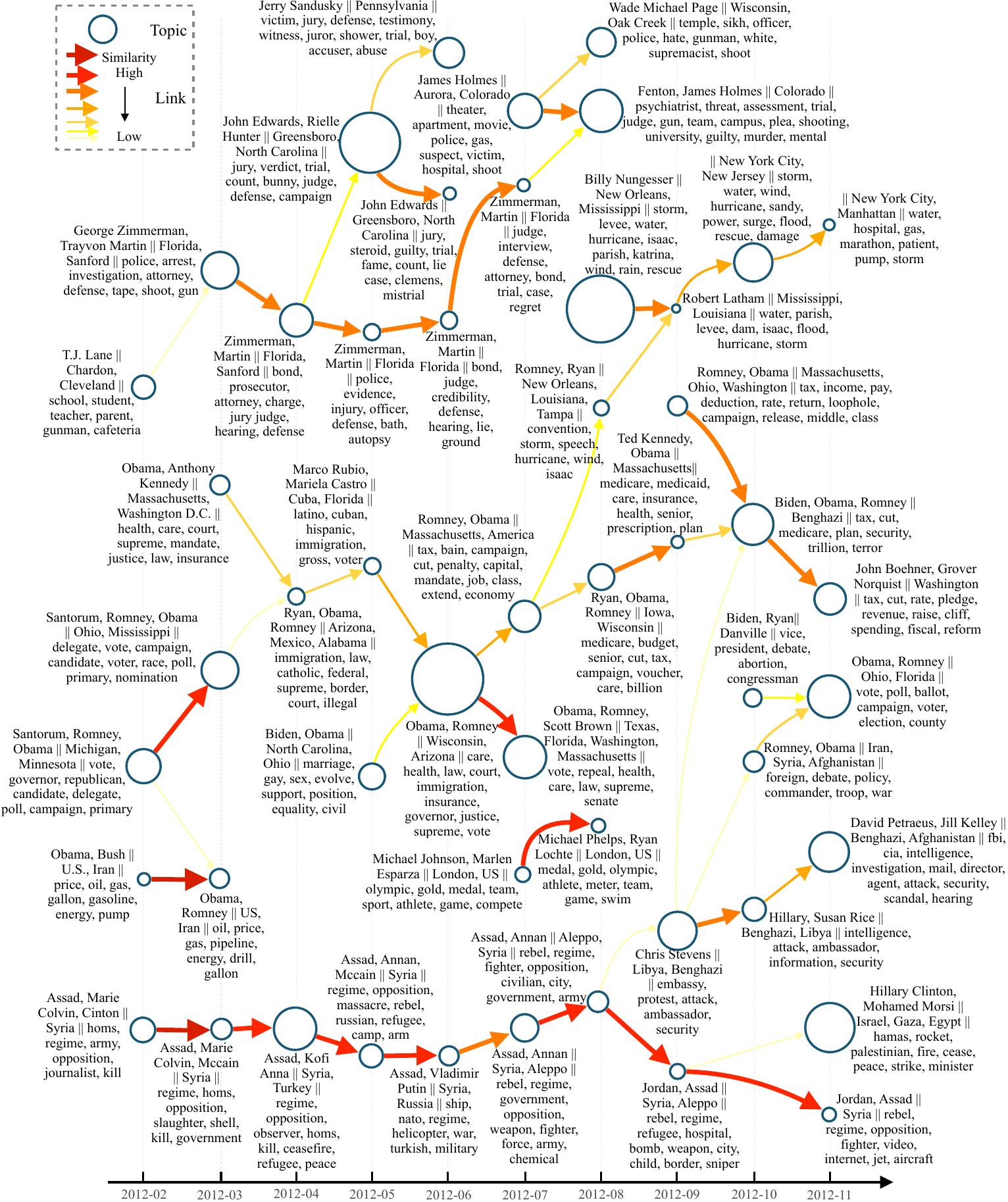}
\caption{Long topic trajectory for 2012 CNN news. Each circle represents one topic and the circle size is proportional to the size of the topic, i.e. the volume of corresponding news stories. Thicker links correspond to greater similarities between topics.}
\label{long_trajectory_2012}
\end{figure*}

To show that our method can work effectively on large-scale datasets, we generate a long topic trajectory using the whole year's news data in 2012 from the network CNN. The obtained trajectory is shown in Fig. \ref{long_trajectory_2012}. Due to the space limit we only show the text part of the topics in the trajectory. The who, where and when parts of the topic are separated by the symbol "$||$" in the figure. The top part of the figure shows the trajectory of George Zimmerman's case and some other related shooting cases such as the Chardon shooting in February and the Colorado theater shooting in July. The middle part of the figure is mainly about topics closely related to the 2012 US election, such as the health care, the immigration problem, the economy and the debates. The Syria problem, which is another factor related to the election, is shown in the bottom part of the figure. We also get some short trajectories such as the one about Olympic shown in the lower half part of the figure. From these trajectories, we can clearly see how these topics develop over time and how they can relate to each other.

We also made a case study based on our method, which tracks the 2016 U.S. presidential election. The large-scale topic detection and tracking results are visualized in the Viz2016 website (mentioned in Section \ref{motivation}).

\section{Conclusions} \label{conclusion}

We have presented a joint image-text news topic detection and tracking method. We use the And-Or graph as a structured topic representation, which models the image and text parts of topics jointly. We detect topics using the SWC-based cluster sampling method. Topics are also tracked over time to deal with the continuous updates of news streams. Qualitative and quantitative evaluation results both show the effectiveness and efficiency of the proposed topic detection and tracking method. 

In the future, we want to expand our study to the problem of media analysis for social and political science research. Based on our topic detection and tracking results, we can analyze how media are biased for different topics, what is the agenda-setting pattern, what is the causal relations between topics, etc.

\section*{Acknowledgment}

This project is supported by the NSF CDI project CNS 1028381. The authors would like to thank Dr. Francis Steen and Tim Groeling at UCLA, and Dr. Chengxiang Zhai at UIUC for discussions and insightful suggestions. We would also like to thank Dr. Quanshi Zhang and Tianfu Wu at UCLA for their assistance.

\ifCLASSOPTIONcaptionsoff
 \newpage
\fi

\bibliographystyle{IEEEtran}
\bibliography{paper}

\end{document}